%% file: ms2.tex
\documentclass[12pt,preprint]{aastex}
\usepackage{amsmath}

\newcommand{\bdv}[1]{\mbox{\boldmath$#1$}}

\def\au{{\rm au}} 
 
\def\kms{{\rm km}\,{\rm s}^{-1}}
\def\masyr{{\rm mas}\,{\rm yr}^{-1}}
\def\kpc{{\rm kpc}}
\def\mas{{\rm mas}}
\def\anom{{\rm anom}}

\def\muas{\mu{\rm as}}

\def\rel{{\rm rel}}

\def\eff{{\rm eff}}
\def\rot{{\rm rot}}

\def\e{{\rm E}}
\def\bpi{{\bdv\pi}}
\def\bmu{{\bdv\mu}}

\def\bgamma{{\bdv\gamma}}

\def\btheta{{\bdv\theta}}

\begin{document}
\title{Mass Production of 2021 KMTNet Microlensing Planets II}

\input author.tex

\begin{abstract}

We continue our program of publishing all planets (and possible planets)
found by eye in 2021 Korea Microlensing Telescope Network (KMTNet) online
data.  We present 4 planets,
(KMT-2021-BLG-0712Lb, 
KMT-2021-BLG-0909Lb, 
KMT-2021-BLG-2478Lb, and 
KMT-2021-BLG-1105Lb), 
with planet-host mass ratios in the range $ -3.3 \la \log q \la -2.2$.
This brings the total of secure, by-eye, 2021 KMTNet planets to 16, including 8
in this series.  The by-eye sample is an important check of the completeness
of semi-automated detections, which are the basis for statistical analyses.
One of the planets, KMT-2021-BLG-1105Lb, is blended with a relatively
bright $(I,V)\sim (18.9,21.6)$ star that may be the host.  This could
be verified immediately by high-resolution imaging.  If so, the host is
an early G dwarf, and the planet could be characterized by radial-velocity
observations on 30m class telescopes.

\end{abstract}


\section{{Introduction}
\label{sec:intro}}
In this paper, we continue the program outlined in Paper I \citep{kb211391}
to ensure the publication of all planets from the Korea Microlensing 
Telescope Network (KMTNet, \citealt{kmtnet}) 2021 season.  As discussed there,
many planets will be published as single-planet papers, either because
of their intrinsic scientific interest or as an entry point of scientific
work by junior workers.  Many others will be published in small groups
that are related by some common thread.  However, robust statistical
investigation requires that all planets be published, or at least be subjected
to publication-quality analysis.  The experience of the 2018 season,
which is the first to be completed 
\citep{kb190253,ob180383,2018prime,2018subprime}, shows that even several years 
after the close of that season, 6 planets that had been detected by eye remained
unpublished, while several dozen other ``possible planets'' required detailed
investigation to determine that they were either non-planetary or ambiguous
in nature.  There were, in addition, 11 planets discovered by the KMT
AnomalyFinder system \citep{ob191053,af2} that had not previously been found
by eye.  As many dozens of KMTNet planets remain to be published from the 2016,
2017, and 2019 seasons, it seems prudent not to fall behind in the publication
of 2021 planets.

As also noted in Paper I, the investigation and publication of all by-eye
discoveries serves as an important check on the AnomalyFinder system.
For 2018, two by-eye discoveries were not recovered by AnomalyFinder:
OGLE-2018-BLG-0677 \citep{ob180677}, which failed to meet the selection
criteria, and KMT-2018-BLG-1996 \citep{kb181976}, which was recovered
in the machine phase of AnomalyFinder but was not finally selected by eye.
Among the $\sim 70$ previously discovered planets from 2016-2019 that met
the selection criteria, KMT-2018-BLG-1996 was one of only two that were not 
recovered.  This was an important check on the completeness of AnomalyFinder.
It is important to maintain this check as the years go forward, and for
this reason, the analysis and publication of 2021 events prior to the
application of AnomalyFinder is crucial to maintaining the robustness of this
check.

In Paper I, we began this process by systematically going through the planetary
candidates that had been selected by YHR, rank ordered by the preliminary
estimates of planet-host mass ratio, $q$.  We published four planets
(KMT-2021-BLG-1391, 
KMT-2021-BLG-1253, 
KMT-2021-BLG-1372, and 
KMT-2021-BLG-0748), 
with finally-adopted mass ratios $-4.4 \la \log q \la -2.9$.

In the present paper, we continue this approach.  We analyze 4
planetary events
(KMT-2021-BLG-0712, 
KMT-2021-BLG-0909, 
KMT-2021-BLG-2478, and 
KMT-2021-BLG-1105). 

From the standpoint of future statistical studies it is just
as important to decisively reject initially plausible candidates
from the final sample as it is to populate the sample.  This
statement is most directly applicable to candidates that are
objectively selected by AnomalyFinder.  However, as there is strong
overlap between by-eye and AnomalyFinder planets, the rejection
of by-eye candidates can contribute substantially to this task.  In Paper I,
we reported that we rejected three such candidates
(KMT-2021-BLG-0637, KMT-2021-BLG-0750, and KMT-2021-BLG-0278).  We note
that in the course of identifying the 4 planetary 
events analyzed here, we rejected 4 others:
KMT-2021-BLG-0631  was eliminated because re-reduction showed that the
apparent anomaly had been due to data artifacts, 
KMT-2021-BLG-0296 and 
KMT-2021-BLG-1484 
were both eliminated because they had low $\Delta\chi^2<30$ improvement relative
to a point lens and (related to this) many competing solutions. In addition,
KMT-2021-BLG-1360 was eliminated because the anomaly detection, although
formally very significant, $\Delta\chi^2=170$, rests on a single point.
One reason for rejecting such ``detections'' is that unexpected systematics
can always corrupt a single point.  Nevertheless, out of intellectual
curiosity, we still conducted a systematic investigation of this event
and found that it had multiple 2L1S solutions that span two decades in $q$,
as well as 1L2S solutions, all at comparable $\chi^2$.  We mention this
mainly as a caution regarding automated planet sensitivity calculations that 
rely solely on $\chi^2$ criteria to determine whether a given simulated planet
is ``detectable''.  In this case, $\Delta\chi^2$ is more than double the
threshold of the KMT AnomalyFinder search algorithm \citep{af2}, yet
the ``planet'' (if that is what caused the anomaly) cannot be recovered
even at order-of-magnitude precision.

Finally, we remark on the progress of publication of other 2021 KMTNet
planets, which, as mentioned above can be of individual or group interest.
In the former category are 
KMT-2021-BLG-0912 \citep{kb210912}, 
KMT-2021-BLG-1077 (two planets) \citep{kb211077},
KMT-2021-BLG-1898 \citep{kb211898}, and 
KMT-2021-BLG-0240 \citep{kb210240}, 
with the last of these probably being unusable for mass-ratio 
studies because of a severe degeneracy in $q$.  In the latter category are the
3 planetary events
KMT-2021-BLG-0320,
KMT-2021-BLG-1303, and
KMT-2021-BLG-1554,
which have the common characteristic of being sub-Jovian planets
\citep{4subjovian}, and 2 others
KMT-2021-BLG-0171, and
KMT-2021-BLG-1689, which have the common characteristic of being discovered
in a survey-plus-followup campaign.  Note that while KMT-2021-BLG-0171 would
have been discovered even without followup data, there are no KMT data during 
the anomaly in KMT-2021-BLG-1689.  Hence, only the first of these two will enter
the AnomalyFinder statistical sample.
  In addition, KMT-2021-BLG-0322 
has been thoroughly investigated and found to be ambiguous between
a binary-star system that may or may not contain a planet \citep{kb210322}. 

Thus, with the publication of this paper, there are a total of about 16
planets from 2021 that are suitable for mass-ratio studies, which
constitutes good initial progress.

\section{{Observations}
\label{sec:obs}}

All of the planets in this paper were identified in by-eye
searches of KMT events that were announced by the KMT AlertFinder
\citep{alertfinder} as the 2021 season progressed.
As described in Paper I, KMTNet observes
from three 1.6m telescopes that are equipped with
$(2^\circ\times 2^\circ)$ cameras at CTIO in Chile (KMTC),
SAAO in South Africa (KMTS), and SSO in Australia (KMTA), mainly
in the $I$ band, with 60 second exposures, but with 9\% of the 
observations in the $V$ band.
The data were reduced using pySIS \citep{albrow09}, a form of
difference image analysis (DIA,\citealt{tomaney96,alard98}).  For publication,
the light curves were re-reduced using the tender-loving care
(TLC) version of pySIS.  
For each event, we manually examined the images during the anomaly
to rule out image artifacts as a potential explanation for the
light-curve deviations.

None of the events reported here were alerted by any other survey, and,
as far as we are aware, there were no follow-up observations.

As in Paper I, Table~\ref{tab:names} gives the event names, 
observational cadences $\Gamma$, discovery dates and sky locations.

\section{{Light Curve Analysis}
\label{sec:anal}}

\begin{figure}
\plotone{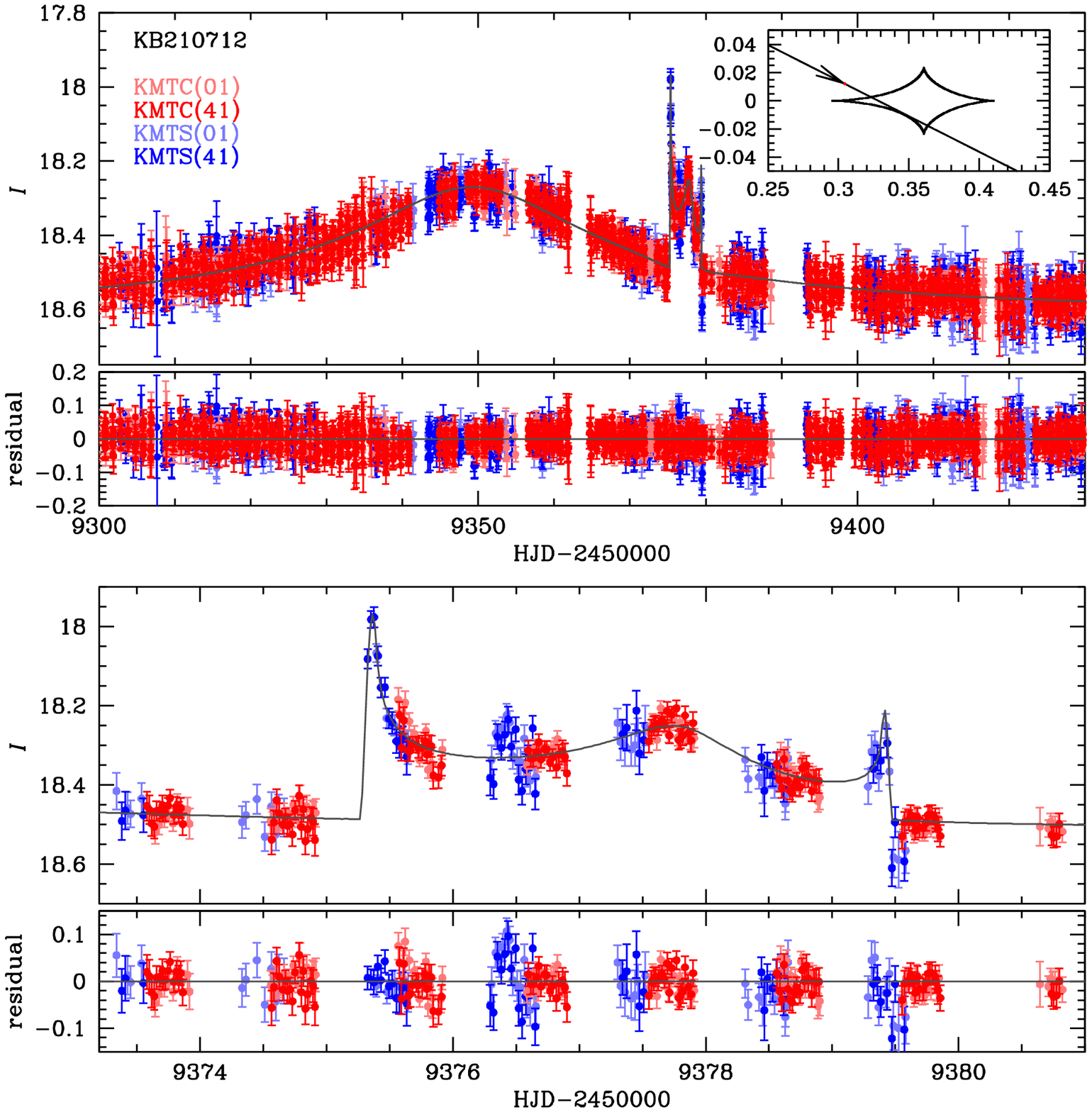}
\caption{Light-curve data and models for KMT-2021-BLG-0712.
Observations are color-coded by observatory and field, as indicated
in the legend.  Residuals to the model are shown for both the
event as a whole (upper) and the anomaly region (lower) panels.
The source trajectory relative to the (major-image planetary) caustic 
is shown in the inset.
}
\label{fig:0712lc}
\end{figure}

{\subsection{{Preamble}
\label{sec:anal-preamble}}

Our approach to analyzing events is identical to that described
in Section~3.1 of Paper I.  Here, we present only the definitions
of the parameter symbols, in conformity with standard practice.
For more details, we refer the reader to Paper I.

All of the events in this paper can be analyzed to a first approximation
as 1L1S events, which are characterized by three \citet{pac86} parameters,
$(t_0,u_0,t_\e)$, i.e., the time of lens-source closest approach, the impact
parameter (normalized to the Einstein radius, $\theta_\e$), and the Einstein
radius crossing time
\begin{equation}
t_\e = {\theta_\e\over\mu_\rel};
\qquad
\theta_\e =\sqrt{\kappa M \pi_\rel},
\qquad
\kappa\equiv {4G\over c^2\,\au}\simeq 8.14{\mas\over M_\odot}.
\label{eqn:thetae}
\end{equation}
Here, $M$ is the mass of the lens, $(\pi_\rel,\bmu_\rel)$ are the lens-source
relative parallax and proper motion, $\mu_\rel\equiv |\bmu_\rel|$,
and $n$L$m$S means ``$n$ lenses and $m$ sources''.

A 2L1S model always requires at least three additional parameters
$(s,q,\alpha)$, i.e., the separation (normalized to $\theta_\e$)
and mass ratio of the two lens components, as well as the angle between
the line connecting these and the direction of $\bmu_\rel$.  
If there are finite-source effects due to the
source approaching or crossing caustic structures that are
generated by the lens, then
one must also specify $\rho\equiv \theta_*/\theta_\e$, where $\theta_*$
is the angular radius of the source.

For 1L2S models, which can generate featureless bumps that can be mistaken
for 2L1S ``planets'' \citep{gaudi98},  the minimal number of parameters is 6, 
including $(t_{0,1},t_{0,2})$ and $(u_{0,1},u_{0,2})$
for the two times of closest approach and impact parameters, respectively,
$t_\e$ for the Einstein timescale, and $q_F$, i.e., the flux ratio
of the two sources in the $I$-band.  In many cases, one or both of the two 
normalized source radii must be specified, $\rho_1=\theta_{*,1}/\theta_\e$ and
$\rho_2=\theta_{*,2}/\theta_\e$.  More complex models involving orbital
motion of the binary-source system may also be needed.

If the microlens parallax effect can be detected (or constrained),
then one should include the microlens parallax vector 
\citep{gould92,gould00,gould04},
\begin{equation}
\bpi_\e = {\pi_\rel\over\theta_\e}\,{\bmu_\rel\over\mu_\rel},
\label{eqn:bpie}
\end{equation}
which is normally expressed in equatorial coordinates
$\bpi_\e=(\pi_{\e,N},\pi_{\e,E})$.  In these cases, one usually must also 
fit, at least initially, for the first derivatives in time of the lens 
angular position, $\bgamma=[(ds/dt)/s,d\alpha/dt]$, because
$\bpi_\e$ and $\bgamma$ can be correlated or even degenerate.
In these cases, we restrict such fits to $\beta<0.8$, where
\citep{eb2k5,ob05071b},
\begin{equation}
\beta\equiv {\kappa M_\odot{\rm yr}^2\over 8\pi^2}{\pi_\e\over\theta_\e}
\gamma^2\biggl({s\over \pi_\e +\pi_s/\theta_\e}\biggr)^3,
\label{eqn:beta}
\end{equation}
and where $\pi_S$ is the source parallax.

In our initial heuristic analyses, we often predict $s^\dagger_\pm$
and $\alpha$ from the morphology of the light curve
\citep{kb190253,kb211391},
\begin{equation}
s^\dagger_\pm = {\sqrt{4 +u_\anom}\pm u_\anom\over 2};
\qquad \tan\alpha = {u_0\over \tau_\anom},
\label{eqn:sdagger}
\end{equation}
under the assumption that the anomaly occurs when the source crosses
the binary axis.  Here, $u_\anom = \sqrt{\tau_\anom^2 + u_0^2}$,
$\tau_\anom = (t_\anom - t_0)/t_\e$, $t_\anom$ is the midpoint of the anomaly,
and the ``$\pm$'' refers to major/minor image perturbations.  If there
are two solutions, with normalized separation values $s_\pm$, 
as often occurs (see \citealt{zhang22} for a theoretical discussion of such 
degeneracies), we expect that the empirical quantity $s^\dagger=\sqrt{s_+ s_-}$
(without subscript)
will be approximately equal to the subscripted quantity from 
Equation~(\ref{eqn:sdagger}).

Finally we often report the ``source self crossing time'',
$t_* \equiv \rho t_\e$.  We note that this is a derived quantity and is
not fit independently.

{\subsection{{KMT-2021-BLG-0712}\label{sec:anal-kb210712}}}

Figure~\ref{fig:0712lc} shows an otherwise standard 1L1S light curve
with \citet{pac86} parameters $(t_0,u_0,t_\e)=(9349.32,0.145,91\,{\rm day})$,
punctuated by a 4.3-day double-horned profile, centered at $t_\anom=9377.35$.
The double-horned profile is unusual in that it has a smooth bump in the 
middle, which is almost certainly generated by the source approaching an
interior wall of the caustic.

{\subsubsection{{Heuristic Analysis}\label{sec:heuristic-kb210712}}}

These parameters imply $\tau_\anom=0.308$, $u_\anom=0.340$, and thus
\begin{equation}
\alpha = 25^\circ;
\qquad
s^\dagger_+ = 1.18.
\label{eqn:kb180712_heur}
\end{equation}
Because the anomaly is clearly due to the source entering
and leaving the caustic, we do not expect a degeneracy in $s$.
Rather, we expect $s\simeq s^\dagger_+$.

{\subsubsection{{Static Analysis}\label{sec:static-kb210712}}}

The grid search on the $(s,q)$ plane returns only one solution,
whose refinement with all parameters set free is shown in 
Table~\ref{tab:0712parms}.  We find that $\alpha$ and $s$ are as expected,
while $\log q=-3.3$ indicates a Saturn mass-ratio planet.  The caustic entrance
and exit are both well-covered, yielding a $\sim 8\%$ measurement
of a relatively low value of $\rho=3.9\times 10^{-4}$.  We will see in
Section~\ref{sec:cmd-kb210712} that this implies a large value of
$\theta_\e\sim 0.64\,\mas$, and so a relatively nearby lens
$\pi_\rel \sim 0.05\,\mas/(M/M_\odot)$ and thus a relatively large
microlens parallax $\pi_\e\sim 0.08\,(\pi_\rel/0.05\mas)$.  Together with the
relatively long timescale and the fact that the anomaly has three
peaks \citep{angould}, this encourages us to search for microlens
parallax solutions, in spite of the relatively faint source, 
$I_{S,\rm KMTC01}\sim 21.6$.

{\subsubsection{{Parallax Analysis}\label{sec:parallax-kb210712}}}

As is almost always the case (except for some extremely long
events), there are two parallax solutions, which are summarized in
Table~\ref{tab:0712parms} and illustrated in Figure~\ref{fig:0712par}.  
As described in Section~\ref{sec:anal-preamble}, we simultaneously fit for the
first derivatives of the planet position due to its orbital motion
$\gamma=[(ds/dt)/s,d\alpha/dt]$.  While Table~\ref{tab:0712parms} gives
$\bpi_\e$ in standard equatorial coordinates, it is also useful to
present these solutions in terms of the principal axes of the error ellipses.
\begin{equation}
(\pi_{\e,\parallel},\pi_{\e,\perp},\psi)
= (+0.123\pm0.028,0.40\pm 0.11,280.4^\circ)\qquad[u_0>0],
\label{eqn:0712par+}
\end{equation}
and
\begin{equation}
(\pi_{\e,\parallel},\pi_{\e,\perp},\psi)
= (+0.033\pm0.028, 0.72\pm 0.14,256.4^\circ)\qquad[u_0<0].
\label{eqn:0712par-}
\end{equation}
Here, $\pi_{\e,\parallel}$ (so called because, for short events, it is
approximately parallel to the projected position of the Sun) is the
minor axis of the error ellipse, $\pi_{\e,\perp}$ is the major axis, and
$\psi$ is the angle of the minor axis, measured north through east.
In line with the sign conventions of Figure~3 of \citet{mb03037}
(keeping in mind that MOA-2003-BLG-037 peaked after opposition while
KMT-2021-BLG-0712 peaked before opposition),
$\pi_{\e,\parallel}$ is approximately west and 
$\pi_{\e,\perp}$ is approximately north.  Note that the actual projected
orientation of Earth relative to the Sun at the peak is $\psi_\odot=281.0^\circ$.
Because the FWHM of the event, $\sqrt{12}u_0t_\e\sim 45\,$days, covers
almost a radian of Earth's orbit, the ``short event'' approximation 
\citep{smp03,gould04,mb03037} is not
expected to yield a precise characterization.}

We find, from fitting the event (with the anomaly removed)
to a point-lens model with parallax, that the presence of the anomaly
reduces the axes of the error ellipses for the two solutions from
(0.030:0.45) to (0.028:0.11) and from (0.032:0.39) to (0.028:0.14),
in particular, reducing the aspect ratios by factors of 3.8 and 2.4
for the two cases.  This confirms the important role of the relatively
complex caustic features in improving the parallax measurement.


{\subsection{{KMT-2021-BLG-0909}\label{sec:anal-kb210909}}}

Figure~\ref{fig:0909lc} shows an otherwise approximately standard 1L1S 
light curve with parameters $(t_0,u_0,t_\e)=(9354.1,0.060,16\,{\rm day})$,
punctuated by a sharp bump, which erupts suddenly at 9360.65 and then
peaks $\Delta t_{\rm rise} =4\,$hr later at $t_\anom=9360.82$.  
This is almost certainly
a caustic entrance, although there is no obvious caustic exit.

{\subsubsection{{Heuristic Analysis}\label{sec:heuristic-kb210909}}}

These parameters imply $\tau_\anom=0.420$, $u_\anom=0.424$, and thus
\begin{equation}
\alpha = 188.1^\circ;
\qquad
s^\dagger_+ = 1.23.
\qquad
s^\dagger_- = 0.81.
\label{eqn:kb180909_heur}
\end{equation}
Note that because the nature of the caustic entrance is unclear,
we report both $s^\dagger_+$ and $s^\dagger_-$.  Moreover, because 
$u_\anom\sim 0.4$ is large, this is a planetary caustic crossing,
so we do not expect a degeneracy in $s$.  Rather, for a major-image
caustic, we expect $s\simeq s^\dagger_+$, while for a minor-image
caustic, we expect a less precise $s\sim s^\dagger_-$ because the caustic
would not lie on the binary axis.

{\subsubsection{{Static Analysis}\label{sec:static-kb210909}}}

The grid search returns only one solution, whose refinement is
described by the parameters given in Table~\ref{tab:0909parms}.
The heuristic estimate of $\alpha$ proves to be too small
by a factor of two relative to the binary axis, i.e.,
$\alpha - 180^\circ = 8.1^\circ$ versus $18.9^\circ$.
This is because the source crosses the binary axis about half way
between the central and planetary caustics, rather than at the planetary
caustic.  See Figure~\ref{fig:0909lc}.  This, in turn, is partly due to the
fact that the planet is relatively massive, $\log q =-2.50$, for which
the caustics are offset from the axis by 
$\eta_{c,-}= 2[q(s^{-2}-1)]^{1/2}\rightarrow 0.078$ \citep{han06}.
See Figure~\ref{fig:0909lc}.  Because the caustic entrance is well-covered
by KMTC data, the normalized source radius, $\rho$, is determined
to better than 10\%.  Given the short Einstein timescale $t_\e=16\,$day
and the faintness of the source, we do not attempt a measurement of $\bpi_\e$.

{\subsection{{KMT-2021-BLG-2478}\label{sec:anal-kb212478}}}

Figure~\ref{fig:2478lc} shows an approximately standard 1L1S 
light curve with parameters $(t_0,u_0,t_\e)=(9482.2,0.08,41\,{\rm day})$,
but with two major features superposed: a poorly sampled caustic feature,
centered at $\sim 9486$, lasting 1.5--2 days, and a roughly 2-day,
roughly symmetric spike, peaking at 9493.9.  The sparse coverage is
primarily due to the fact that these anomalies occurred near the
end of the microlensing season, when the field was visible only about
3 hours per night from each site, and partly due to episodes of adverse
weather.

The first (i.e., caustic) structure implies that there must be a second lens.
If the system is not more complicated than this, i.e., it is 2L1S,
then the presence of two anomalies at $\tau_1\sim +0.09$ and
$\tau_2\sim +0.29$ after peak almost certainly implies a very large
resonant caustic.  In principle, however, such multiple
anomalies might require more complex systems, such as 3L1S.

{\subsubsection{{Heuristic Analysis}\label{sec:heuristic-kb212478}}}

Within the 2L1S framework, $t_\anom = 9493.9$, i.e., $\tau_\anom=\tau_2=0.29$,
and $u_\anom=0.30$, so
\begin{equation}
\alpha = 14.9^\circ;
\qquad
s^\dagger_+ = 1.16 .
\label{eqn:kb182478_heur}
\end{equation}

{\subsubsection{{Static Analysis}\label{sec:static-kb212478}}}

The grid search yields only one solution, whose refined parameters
are given in Table~\ref{tab:2478parms}.  Note that while the
heuristic $\alpha$ prediction was approximately correct, the
heuristic $s^\dagger_+$ (combined with $s=1.058$ from 
Table~\ref{tab:2478parms}), predicts a second solution at 
$s_{\rm inner} = (s^\dagger_+)^2/s_{\rm outer}=1.27$.  Such solutions can
generate a cusp-approach spike as the source passes over the ridge
between the central and planetary caustics, but the central caustic
is not large enough to induce the first caustic anomaly that is
seen in the light curve.  Hence, there is no degeneracy.

As with KMT-2021-BLG-0909, this planet has a super-Jovian mass ratio.

The Markov Chain Monte Carlo (MCMC) 
constraints on $\rho$ are not adequately summarized by
the median and 68 percentile format of Table~\ref{tab:2478parms}.
The main takeaway would be that, at the $1\,\sigma$ level, 
$\rho<1\times 10^{-4}$.  In Section~\ref{sec:cmd-kb212478}, we will show that
$\theta_*\sim 0.5\,\muas$.  Hence, this limit would imply $\theta_\e> 5\,\mas$,
which would be quite extraordinary.
We defer investigation of the reliability of such limits to 
Section~\ref{sec:parallax-kb212478}, but for the moment we simply note that
we must at least consider the possibility of very large 
$\theta_\e =\sqrt{\kappa M \pi_\rel}$ and so very nearby and/or very massive
lenses.  The former would imply a large, hence potentially measurable 
microlens parallax $\pi_\e$.  As mentioned in Section~\ref{sec:static-kb210712},
the complex caustic structure, spanning a significant fraction of $t_\e$, 
greatly enhances the prospects for making such a measurement.

{\subsubsection{{Parallax Analysis}\label{sec:parallax-kb212478}}}

Including $\bpi_\e$ and $\bgamma$ improves the fit by $\Delta\chi^2=146$,
with the $u_0>0$ and $u_0<0$ being almost perfectly degenerate.  
See Table~\ref{tab:2478parms}.  The most important aspect of these
fits is that $\pi_\e\simeq \pi_{\e,E}=0.52\pm 0.07$ 
(because $|\pi_{\e,E}|\gg |\pi_{\e,N}|$) is indeed large.

However, these solutions imply that the $\rho$
measurement requires closer examination.  In particular, the $1\,\sigma$
limit remains similar, which, if accepted at face value, would
imply $M>1.2\,M_\odot$ and $D_L<0.4\,\kpc$.  Clearly, such a lens would
be so bright, $I_L\la 12$, 
as to prevent microlensing observations in its neighborhood,
unless it were a black hole or neutron star.

Hence, we must investigate the origins of the $\rho$ limit in the light curve
and also consider the extent to which it can be relaxed within acceptable
statistical limits.

The constraints on $\rho$ come entirely from the curvature in the KMTA
data during the spike, which drop by $\sim 0.1\,$mag over the course
of $\delta t\sim 0.25\,$hr.  Because the source is crossing the ridge
extending from the caustic at a steep angle 
$\alpha^\prime=\alpha+ (d\alpha/dt)\Delta t_2 = 16^\circ$ 
(where $\Delta t_2=11.1\,$day), the intrinsic timescale of this feature
is foreshortened to $\delta t\sin\alpha^\prime= 0.07\,$hr, which is driving
the extremely short $t_*<0.08\,$hr in Table~\ref{tab:2478parms}.

These late-season, end-of-night data were taken at 
high (for KMT) airmass of $\sim 1.9$, which raises the possibility
of a spurious decline in flux due to deteriorating seeing or other effects.
Indeed, we find that the structure of the KMTA peak is remarkably well
anti-correlated with seeing over the whole night.  However, after conducting
several tests, we find no evidence for flux-seeing correlations in other
parts of the KMTA light curve.  Thus, we cannot simply reject this light-curve
structure as seeing-induced.

Next, we investigate in more detail the statistical limits on $\rho$ 
under the assumption that systematics play no role.  We carry out 
fits including $\bpi_\e$ and $\bgamma$ with $\rho$ fixed at various values.
We find that values of $\rho=(1,2,4,8)\times 10^{-4}$ are disfavored at
$\Delta\chi^2 =(1.4,3.2,7.4,19.4)$.  Thus, we regard $\rho=3\times 10^{-4}$
as marginally acceptable even assuming Gaussian statistics.  If we further
take account of possible systematics from end-of-night data taken at
high airmass, even if we cannot identify a specific physical cause,
the constraints become weaker.  Therefore, we will treat the $\rho$
limits cautiously when we investigate the physical nature of the system
in Section~\ref{sec:phys-kb212478}.

{\subsection{{KMT-2021-BLG-1105}\label{sec:anal-kb211105}}}

Figure~\ref{fig:1105lc} shows an otherwise standard 1L1S light curve
with parameters $(t_0,u_0,t_\e)=(9375.8,0.11,35\,{\rm day})$,
punctuated by a sharp spike at $t_\anom = 9373.4$, i.e., 2.3 days before peak.

{\subsubsection{{Heuristic Analysis}\label{sec:heuristic-kb211105}}}
These parameters imply $\tau_\anom = 0.07$, $u_\anom = 0.13$, and so
\begin{equation}
\alpha = 122^\circ;
\qquad
s^\dagger_+ = 1.067 .
\label{eqn:kb1105_heur}
\end{equation}

{\subsubsection{{Static Analysis}\label{sec:static-kb211105}}}

Somewhat surprisingly, the grid search returns 6 local minima.  After 
refinement, we reject two of these because they have high
$\Delta\chi^2=69$ and 109 and, moreover, have poor fits by eye.
However, we briefly note that both have relatively high mass ratios
$\log q \sim -1.75$ and for both, the spike arises from an off-axis
cusp approach to a resonant caustic.  While these models are certainly
not correct, they emphasize the importance of making a systematic search of
parameter space because the overall appearance of the models is not
qualitatively different from the observed light curve.

The remaining four models are shown in Figure~\ref{fig:1105lc}, with
the corresponding geometries shown in Figure~\ref{fig:1105caust},
while their refined parameters are given in Table~\ref{tab:1105parms}.
Locals 1 and 2 constitute an inner/outer degeneracy with 
$s^\dagger=1.068$, while Locals 3 and 4 constitute a second 
inner/outer degeneracy with $s^\dagger=1.066$, both in excellent agreement
with Equation~(\ref{eqn:kb1105_heur}).  This again emphasizes the
importance of a systematic search.  Because Locals 3 and 4
are each disfavored by $\Delta\chi^2>10$, we consider that these solutions
are excluded. Nevertheless, it is notable that these two pairs of solutions
differ in $q$ by more than a factor of 2.

This is another super-Jovian mass-ratio planet, $\log q=-2.7$.

We note that while $\rho$ is not measured, the constraint, $\rho<0.0013$,
at $2.5\,\sigma$, corresponding to $t_*< 1.1\,$hr,
is strong enough to play a significant role.  That is, in 
Section~\ref{sec:cmd-kb211105}, we will show that $\theta_*=0.5\,\muas$,
implying $\mu_\rel = \theta_*/t_* > 4\,\masyr$, which excludes a significant
part of proper-motion parameter space.  Hence, when we incorporate the
$\rho$ constraint to estimate the physical parameters of the system
in Section~\ref{sec:phys-kb211105}, we apply the full $\chi^2(\rho)$
envelope function, rather than a simple limit.  For the moment, we simply
note that even the $1\,\sigma$ ``limit'' corresponds to $\mu_\rel > 6.5\,\masyr$,
and so still leaves a substantial range of values that are well-populated
by Galactic models.  This fact will become relevant in 
Section~\ref{sec:binary-kb211105}.

Due to the faintness of the source and the lack of complex anomaly
structures, we do not attempt a parallax analysis.

{\subsubsection{{Binary-Source Analysis}\label{sec:binary-kb211105}}}

As with all bump-like anomalies that lack complex or caustic-crossing
features, we must check whether the anomaly can be produced by 
a second source (1L2S) rather than a second lens (2L1S).  The results
are shown in Table~\ref{tab:1105_1L2S}.

There are two main features to note about this solution.  First,
while the $\chi^2$ difference,
$\Delta\chi^2 = \chi^2({\rm 1L2S}) - \chi^2({\rm 2L1S}) = 5.5$,
favors the 2L1S solution, it is not large enough, by itself,
to definitively rule out the 1L2S solution.

Second, the value of the second-source self-crossing time,
$t_{*,2} = 1.28\,$hr, is well-measured in this model (contrary to 2L1S),
with just a 5\% error.  At first sight, this value appears to be
very ``typical'' of historic measurements of $t_*$ for dwarf-star sources
in microlensing events.  However, in this instance, the source is
about 100 times fainter than typical cases, $I_{S,2}=26.3$.  We will
show in Section~\ref{sec:cmd-kb211105} that this implies 
$\theta_{*,2}=0.169\,\muas$ and thus $\mu_\rel=\theta_{*,2}/t_{*,2}=1.16\,\masyr$.
Only a fraction of $p<(\mu_\rel/\sigma_\mu)^3/6\sqrt{\pi}\rightarrow 0.006$
microlensing events will have such low proper motions 
(e.g., \citealt{gould21,2018prime}).  Here, we have approximated the bulge
proper-motion distribution as an isotropic Gaussian, with 
$\sigma_\mu = 2.9\,\masyr$.  For example, in a systematic study of 30
1L1S events with finite-source effects (thus permitting $\mu_\rel$ measurements),
which was sensitive to $\mu_\rel \geq 1.0\,\masyr$, \citet{fspl2} found that
the slowest (KMT-2019-BLG-0527) had $\mu_\rel=1.45\,\masyr$, i.e.,
$\sigma_\mu/2$.  See their Figure~5.  Thus, the combination of the
$\Delta\chi^2$ preference discussed above, together with this
kinematic argument, overwhelmingly favors the 2L1S (i.e., planetary)
interpretation.

For completeness, we remark that because the second source would be
very red, the 1L2S model predicts that the bump-anomaly would be
much less pronounced in the $V$ band than the $I$ band.  See, e.g.,
\citet{ob151459} for a practical example.  Unfortunately, however,
there are no $V$-band data during the anomaly.

Because the interpretation of the event rests heavily on the kinematic
argument, we must also consider the possibility that this argument can be evaded
(at some cost in $\chi^2$) by solutions with much smaller $\rho$.
We first check that the $1\,\sigma$ error bar on $\rho_2$ shown in
Table~\ref{tab:1105_1L2S} is actually representative of the $\chi^2$
surface out to $3\,\sigma$ by fixing $\rho_2$ at various values.
We find that it is.  See Table~\ref{tab:1105_1L2S} for an example.
Next, we search for solutions that are away from
this local minimum by enforcing $\rho_2=0$.  We find that there is
such a solution, but it is disfavored by $\Delta\chi^2=13.4$.  
See Table~\ref{tab:1105_1L2S}.  Thus,
while this solution avoids the proper-motion constraint, it
increases the total $\chi^2$ difference to
$\Delta\chi^2 = \chi^2({\rm 1L2S}) - \chi^2({\rm 2L1S}) = 19$.
This would be high enough to decisively reject 1L2S were
we to adopt the low-$\rho_2$ solution.

Moreover, there is an additional statistical argument against the
1L2S solution,  From the Local-1 panel of
Figure~\ref{fig:1105caust}, it is clear that there
is a range of ``$x$'', i.e., $u_x$, of about 0.15 Einstein radii that
would generate a qualitatively similar non-caustic-crossing bump.  However,
the 1L2S solution requires the source to cross the face of the second source,
which has a probability of $p=2\rho_2\simeq 0.003$, i.e., about 50 times
smaller.  To fully evaluate this relative probability, we would have
to consider the relative probabilities of the presence of lens planetary 
companion, compared to a source M-dwarf companion, which we do not attempt here
because it is unnecessary to make the basic argument.  Nevertheless, it is clear
that the 1L2S solution requires some fine tuning.

While we cannot absolutely rule out the 1L2S solution, the formal probability
that it is correct is about $p\sim 4\times 10^{-4}$.  Hence, this
planet should be accepted as genuine.  We note that its reality can
be definitively tested at first adaptive optics (AO) light on next generation
(30m) telescopes, roughly in 2030, i.e., $\Delta t=9\,$yr after the event.
If, as anticipated, the 2L1S model is correct, then the source and lens 
will be separated by $\Delta\theta = \mu_\rel\Delta t\ga 36\,\mas$, so they
will be easily resolved.  On the other hand, if the 1L2S model were correct,
then the separation would be $\Delta\theta \sim 9\,\masyr$, which would 
probably be too small to resolve, but even if resolved would provide a
measurement that was consistent with 1L2S but not with 2L1S.

Before leaving the issue of 1L2S models, we note that as a matter
of ``due diligence'', we explored 1L2S models in which finite-source
effects were permitted for both the primary and secondary sources,
even though such effects are extremely unlikely for the primary, 
a priori, because
$t_\eff\equiv u_0 t_\e\sim 4\,$days is extremely long relative to the
typical self-crossing time of dwarf stars, $t_* \sim 1\,$hr.  Surprisingly,
we did indeed find such solutions with $\Delta\chi^2\sim -6$ relative
to the solution reported in Table~\ref{tab:1105_1L2S} (so, comparable
$\chi^2$ to the 2L1S solution).  However, these had 
$(\rho_1,\rho_2)\simeq (0.2,0.002)$, which would imply grossly inconsistent
estimates for $\theta_\e=\theta_*/\rho$ of $2.5\,\muas$ versus $85\,\muas$.
Hence, it is unphysical.  The $\chi^2$ improvement could be a purely
statistical fluctuation ($p=0.05$) or it could be due to low level
systematics in the photometry.  In any case, we reject this solution.

Finally, we remark that this event was included in the present study
only because our ``mass production'' project aims to document all
2021 events with viable planetary solutions, in the spirit pioneered
by \citet{2018prime} and \citet{2018subprime} for 2018 events, 
irrespective of whether such planetary solutions are decisively preferred.
Our initial assessment, based on detailed modeling of TLC reductions,
was that its interpretation was ambiguous, and thus it would not
enter planetary catalogs.  It was only in the course of comprehensively
evaluating all the evidence that we concluded that the planetary solution
is decisively favored.





\section{{Source Properties}
\label{sec:cmd}}

Our evaluation of the source properties exactly follows the goals
and procedures of Paper I.  In this introduction, we repeat only
the most essential descriptions from Section~4 of that work,
in particular (as in Section~\ref{sec:anal-preamble}) documenting
all notation.

We analyze the color-magnitude diagram (CMD) of each
event, primarily to measure $\theta_*$ and so to determine
\begin{equation}
\theta_\e = {\theta_*\over\rho};\
\qquad
\mu_\rel = {\theta_\e\over t_\e}.
\label{eqn:thetae_murel}
\end{equation}
We follow the method of \citet{ob03262}.  We first find the offset
of the source from the red clump
\begin{equation}
\Delta[(V-I),I] = [(V-I),I]_S - [(V-I),I]_{\rm cl}.
\label{eqn:deltacmd}
\end{equation}
We adopt  $(V-I)_{\rm cl,0}=1.06$ from \citet{bensby13} and evaluate
$I_{\rm cl,0}$ from Table~1 of \citet{nataf13}, based on the Galactic
longitude of the event, which yields the  dereddened color and magnitude
of the source,
\begin{equation}
[(V-I),I]_{S,0} = [(V-I),I]_{\rm cl,0} + \Delta[(V-I),I].
\label{eqn:cmd0}
\end{equation}
Next, we transform from $V/I$ to $V/K$ using the $VIK$ color-color 
relations of \citet{bb88}, and we apply the color/surface-brightness relations
of \citet{kervella04} to obtain $\theta_*$.  After propagating the
measurement errors, we add 5\% to the error in quadrature to take account
of systematic errors due to the method as a whole.

To obtain $[(V-I),I]_S$, we always begin with pyDIA reductions \citep{pydia},
which put the light curve and field-star photometry on the same system.
With one exception (see below), we determine $(V-I)_S$ by regression of the 
$V$-band data 
on the $I$-band data, and we determine $I_S$ by regression of the $I$-band
data on the best-fit model.  For 2 of the 4 events analyzed in this
paper, there is calibrated OGLE-III field-star photometry \citep{oiiicat}.
For these 2 cases, we transform $[(V-I),I]_S$ to the OGLE-III
system.  For the 2 remaining cases, we work in the instrumental 
KMT pyDIA system.

For KMT-2021-BLG-0909, the source is too faint in the $V$ band to measure
the source color from the light curve.  We therefore employ a different
technique, as described in Section~\ref{sec:cmd-kb210909}.

The CMDs are shown in Figure~\ref{fig:allcmd}.

The elements of these calculations are summarized in
Table~\ref{tab:cmd}.   
In all cases, the source
flux is that of the best solution.  Under the assumption of fixed
source color, $\theta_*$ scales as $10^{-\Delta I_S/5}$ for the other solutions,
where $\Delta I_S$ is the difference in source magnitudes, as given
in the Tables of Section~\ref{sec:anal}.
The inferred values (or limits
upon) $\theta_\e$, and $\mu_\rel$ are given in the 
individual events subsections below, where we also discuss other
issues, when relevant.

{\subsection{{KMT-2021-BLG-0712}\label{sec:cmd-kb210712}}}

There are two issues related to the source that require some care for this
event.  First, the source color shown in Table~\ref{tab:cmd}, 
$(V-I)_{S,0}=0.69\pm 0.06$,
is unusually blue given that the source lies $\Delta I=5.9$ magnitudes
below the clump.  If the source were a typical bulge star of this
brightness, we would expect $(V-I)_{S,0}\sim 1.1$, based on
{\it Hubble Space Telescope} images of Baade's Window taken by
\citet{holtzman98}.  Logically, there are three possibilities:
our color measurement is incorrect; the source lies well behind the bulge
and thus is much more luminous (and so bluer) than a bulge star of similar
brightness; or the source is atypical, e.g., has much lower metallicity
than typical bulge stars.  The first of these explanations is the only
one of direct concern here: if the color and magnitude of the source
are correctly measured, regardless of the exact cause of it being so
blue, then the derived $\theta_*$ will also be correct.

We therefore check the color determination as follows:
The color and magnitude reported in Table~\ref{tab:cmd}
are based on the KMTC41 data set.  We repeat the calculation using
the KMTC01 data set, which is composed of a completely independent
series of observations.  While these observations are made with the
same (KMTC) telescope, the observational times are different, and
the positions on the focal plane are offset by $8^\prime$.  Yet, the
best-fit color is the same to within 0.01 mag.  Neither of the other
two explanations appear likely a priori.  To be sufficiently more
luminous to account for the color discrepancy, the source should be
roughly a factor of 2 more distant than the bulge, which would
place it almost 1 kpc below the Galactic plane.  While there are
certainly some stars at this height and this Galactocentric
radius (i.e., similar to that of the Sun), they are relatively rare.
Extremely metal-poor stars in the bulge are likewise rare. 

Despite the low prior likelihood of either of these two options,
they are not unphysical, and hence we adopt the measured color,
and so the value of $\theta_*=0.255\,\muas$ given in Table~\ref{tab:cmd},
and we thereby derive,
\begin{equation}
\theta_\e = 0.604\pm 0.095\,\mas;
\qquad
\mu_\rel = 2.19\pm 0.34\,\masyr,
\qquad (u_0<0)
\label{eqn:0712thetae_minus}
\end{equation}
and
\begin{equation}
\theta_\e = 0.636\pm 0.091\,\mas;
\qquad
\mu_\rel = 2.62\pm 0.37\,\masyr,
\qquad (u_0>0)
\label{eqn:0712thetae_plus}
\end{equation}

The second issue that requires some care is the location of the
blend relative to the source.  If these were closely aligned,
it would argue for the blend being associated with the event,
either being the lens itself or a companion to the lens or the source.

In the KMTC41 pyDIA analysis, the baseline object appears to lie
$\Delta\btheta(N,E) = (170,145)\,\mas$ northeast of the source.
The issue that requires care is that there is another, slightly
brighter star that lies $1^{\prime\prime}$ northwest of the baseline
object, which could in principle corrupt the astrometry of the
baseline object.  (The position of the source is determined
from difference images, for which no such issues arise.)\ \
We conduct two tests.  First, we repeat the analysis using the KMTC01
observations and find almost exactly the same result.  Second,
we find, after transforming coordinates to the OGLE-III system,
that the offset is qualitatively similar: 
$\Delta\btheta(N,E) = (80,250)\,\mas$.  Note that because the epoch of
the OGLE-III data is 15 years earlier, we expect offsets of order 50 mas
in each direction, in addition to normal measurement errors.
The blend is 0.12 mag bluer and and 3.38 mag fainter than the clump
(see Figure~\ref{fig:allcmd}), and it is therefore likely to be
a bulge subgiant.  We conclude that it is most likely not related to the event.
The lens must be fainter than the blend, but 
because the two are separated by just 220 mas, we cannot place more
stringent constraints on the lens light than this.

{\subsection{{KMT-2021-BLG-0909}\label{sec:cmd-kb210909}}}

Due to high extinction, $A_I\sim 4$, the source is too faint in the $V$ band
to measure the source color from the light curve.  In such cases, one
generally estimates the source color based on its offset in the $I$ band
from the centroid of the red clump.  As often happens for such heavily
reddened fields, it is difficult to precisely locate the red clump
on the pyDIA (or, when available, OGLE-III) CMD because even red clump
stars are near or below the measurement threshold in the $V$ band.
In the present case, we find that red clump is detectable on the pyDIA
CMD, but its centroid cannot be reliably determined because the lower
part of the clump merges into the background noise of the diagram.

Therefore, we measure the clump centroid on an $[(I-K),I]$ CMD, which we
construct by matching pyDIA $I$-band photometry with $K$-band photometry from
the VVV catalog \citep{vvvcat}.  See Figure~\ref{fig:allcmd}.  To estimate
the color, we first find the offset from the clump 
$\Delta I=I_S - I_{\rm cl}=2.85\pm 0.08$.  See Table~\ref{tab:cmd}.
If the source were exactly
at the mean distance of the clump, it would therefore have an absolute
magnitude, $M_I=2.73$.  In fact, it is more likely to be toward the back
of the bulge (because it must be behind the lens), so a plausible range
of possibilities is $2.2 \la M_I \la 2.9$.  In this range, the source
could be almost anywhere along the turnoff/subgiant branch.  To account for
this, we adopt a uniform distribution, $0.60<(V-I)_{S,0}<1.00$, which we 
summarize as a $1\,\sigma$ range of $(V-I)_{S,0} = 0.80\pm 0.12$.
This source position is illustrated in Figure~\ref{fig:allcmd} by
transforming from $(V-I)$ to $(I-K)$ using the relations of \citet{bb88}.
These values lead to estimates of
\begin{equation}
\theta_\e = 0.362\pm 0.049\,\mas;
\qquad
\mu_\rel = 8.24\pm 1.11\,\masyr.
\label{eqn:0909thetae}
\end{equation}

Also shown in the CMD is the position of the blended light, which
is a bright giant that is more than 1 mag above the clump.  We find
that this star is displaced by $0.73^{\prime\prime}$ from the source toward
the southwest.  This bright star is almost certainly not associated with
the event, but it prevents us from placing any useful limits on the lens light.

For completeness, we note that the coordinates shown for this
event in Table~\ref{tab:names} are, as usual, those of the nearest catalog
star, namely the bright giant just discussed.  However, these differ
from the coordinates shown on the KMT webpage, which are about 
$1.5^{\prime\prime}$ yet farther south.  When the event was originally triggered
by AlertFinder \citep{alertfinder}, it was identified with this more southerly
catalog star.  One day later, it was again triggered, this time 
by the closer (bright)
catalog star, but our standard procedures enforce maintaining the coordinates
of the original announcement on the web page to avoid confusion.

{\subsection{{KMT-2021-BLG-2478}\label{sec:cmd-kb212478}}}

The source star, whose parameters are given in Table~\ref{tab:cmd} and
whose CMD position is shown in Figure~\ref{fig:allcmd}, lies 4.4 mag
below the clump and is about $0.07\pm 0.06$ mag redder than the Sun.
That is, it is a bulge middle-G dwarf.  Unfortunately, as discussed
in Section~\ref{sec:anal-kb212478}, we have only a $\chi^2$-envelope
constraint on $\rho$, rather than a measurement.  
See Figure~\ref{fig:2478envelope}.  For the present, we
therefore present the estimates of $\theta_\e$ and $\mu_\rel$ scaled
to that section's ``marginally acceptable limit'' (assuming Gaussian errors),
\begin{equation}
\theta_\e = {3\times 10^{-4}\over \rho} (1.78\pm 0.014)\,\mas
\qquad
\mu_\rel = {3\times 10^{-4}\over \rho}(19.3\pm 1.6)\,\masyr.
\label{eqn:2478thetae}
\end{equation}
The high value of $\mu_\rel$ is particularly unexpected.  If there were
no reasons to suspect that this might be due to systematics in end-of-night
data, it would have to be ``cautiously accepted''.  In our actual case,
it invites serious doubt on the reliability of the $\rho$ measurement.

A more robust kinematic constraint comes from $\pi_\e=0.53\pm 0.07$, which
yields a projected velocity of
\begin{equation}
\tilde v \equiv {\au\over \pi_\e t_\e} = 97\pm 13\,\kms.
\label{eqn:2478tildev}
\end{equation}
From this, one may make a rough estimate of the lens distance
(e.g., \citealt{han95}), 
$\pi_\rel = (v_\rot/\tilde v)(\au/R_0)\rightarrow 0.30\,\mas$, i.e.,
$D_L\sim 2.4\,\kpc$,
where $v_\rot \simeq 235\,\kms$ is the rotation speed of the Galaxy and
$R_0\simeq 8\,\kpc$ is the Galactocentric distance.  Because this
rough estimate is in only mild tension with the ``surprising'' result
in Equation~(\ref{eqn:2478thetae}), we will, 
in Section~\ref{sec:phys-kb212478}, consider and compare results
that both include and remove the constraints on $\rho$.

Figure~\ref{fig:allcmd} shows the location of the blended light,
which is 2.2 mag brighter than the source and of similar color.
We find that the source is displaced from the baseline object by
$\Delta\theta=220\,\mas$ to the southeast.  It is consistent with
being a bulge turnoff/subgiant star and thus could in principle
be a companion to the source, but there is no strong evidence
in favor of this hypothesis.
In Section~\ref{sec:phys-kb212478}, we will impose the constraint
on lens light: $I_L>I_B$.  This constraint will further support the
caution regarding the $\rho$ measurement.
For example, if the lens has $\pi_\rel=0.30\,\mas$
as crudely estimated above from the kinematic argument (which ignores the
$\rho$ constraint), then $M\simeq 0.12\,M_\odot$ and 
$D_L\simeq 2.4\,\kpc$, which would be far fainter than this limit on lens light.
However, if we were to accept the ``marginally acceptable limit'' on $\rho$,
then $M>0.4\,M_\odot$ and $D_L<1\,\kpc$, which would imply that the lens
light exceeds this limit, unless the lens were a remnant.

{\subsection{{KMT-2021-BLG-1105}\label{sec:cmd-kb211105}}}

As shown in Table~\ref{tab:cmd}, the source star lies 4.24 mag below
the clump and is measured to have $(V-I)_{S,0} = 0.51\pm 0.09$.  This
is unexpectedly blue, although it is within $1\,\sigma$ of a plausible
value for a relatively metal-poor turn-off star.  We attempt to check
this measurement using KMTS data.  However, these have too few magnified
$V$-band points for a reliable measurement.  We adopt the orientation that
our normal error estimates adequately cover the measurement uncertainty.
As discussed in Section~\ref{sec:anal-kb211105}, we obtain only
a $\chi^2(\rho)$ envelope function which we will incorporate
into the Bayesian analysis in Section~\ref{sec:phys-kb211105}.
See Figure~\ref{fig:1105envelope}.
Hence, the $\theta_*$ determination in Table~\ref{tab:cmd} does not
lead to unambiguous estimates of $\theta_\e$ and $\mu_\rel$.  These
can be fully investigated only in the context of the Bayesian analysis.
For the moment, we express them in parametric form,
\begin{equation}
\theta_\e = {8.2\times 10^{-4}\over\rho }\times (0.59\pm 0.07)\,\mas;
\qquad
 \mu_\rel = {8.2\times 10^{-4}\over\rho }\times (6.1\pm 0.7)\,\masyr,
\label{eqn:thetae1105}
\end{equation}
where the prefactor has values $\simeq (1,0.71,0.57)$ at 
$\Delta\chi^2=(1,4,9)$ of the envelope function.  Thus, in contrast
to many cases that lack a clear $\rho$ measurement, the $\rho$
constraint will play a significant role.

As also discussed in Section~\ref{sec:anal-kb211105}, evaluating the angular
radius of the second source in the 1L2S solution, $\theta_{*,2}$, is
critically important to the kinematic argument against this solution.
We present a new method for doing so, which is particularly adapted
to mid-late M dwarfs, for which it may be very difficult to make the color
measurements that are needed for the traditional \citep{ob03262} method.  The 
first step is to note, from Tables~\ref{tab:1105parms} and \ref{tab:1105_1L2S},
that this second source is 5.11 mag fainter than the source in the 
Local-1 2L1S model, which (from Table~\ref{tab:cmd}) is 4.22 mag fainter than
the clump.  That is, the second source is $9.33\pm 0.11$ mag fainter than the 
clump, where the error is the quadrature sum of the errors in $I_{S,2}$ 
(Table~\ref{tab:1105_1L2S}) and $I_{\rm cl}$ (Table~\ref{tab:cmd}).
We adopt $I_{\rm cl,0}=14.37$ and  $M_{I,\rm cl}=-0.12$.

We do not know, a priori, the exact distance of the source system along the line
of sight.  As noted above, the primary source appears to be a bulge turnoff star
and so could in principle be anywhere in the bulge.  We will consider below
the full range of distances, but for the moment we adopt a fiducial distance
modulus ${\rm Dmod}_{\rm fid} = 14.37 - (-0.12) + 0.2 = 14.69$, i.e., 
0.2 mag behind the clump centroid.  

Next, we evaluate the second source radius under the assumption that it lies
exactly at this distance, and we initially ignore the error in its flux
measurement.  That is, we initially assume that it has an absolute
magnitude $M_{I,2} = 9.33 + (-0.12) -0.20 = 9.01$.
Using the mass-luminosity relations of
\citet{benedict16} in $V$ and $K$, together with the $VIK$ color-color 
relations of \citet{bb88}, we find that the mass of the putative second 
source would be $M_{S,2} = 0.314\,M_\odot$.   We then adopt the
M-dwarf mass-radius relation $(R/R_\odot)=(M/M_\odot)$ from Figure~7 of
\citet{parsons18}, and so obtain $R_2 = 0.314\,R_\odot$ and thus
$\theta_{*,2} = 0.169\,\muas$.

We now take account of the fact that the source system could be at
other distance moduli in the bulge.  For example, if it were at 0.1
mag larger Dmod, then it would likewise be 0.1 mag more luminous,
and so would have correspondingly larger mass (and radius), but the
impact of this larger physical size on $\theta_{*,2}$ would be countered
by the larger distance of the system.  Applying the above arguments to
arbitrary distances (within the bulge), we find,
$d M_I/d\ln M = -2.430$ and so (after a few steps),
\begin{equation}
\theta_{*,2} = 0.169\,\muas \times 10^{-0.0213\,\Delta{\rm Dmod}},
\label{eqn:2478thetastar2}
\end{equation}
where $\Delta{\rm Dmod}$ is the difference between the true distance modulus
and the fiducial one adopted above.  Stated alternatively, 
$\theta_{*,2}\propto D_S^{0.046}$.  Thus, for example, if we adopt
$\Delta{\rm Dmod} = 0\pm 0.3$, then (still not taking account of the 
measurement error of $\sigma(I_{s,2})=0.11$), 
we find $\theta_{*,2} = 0.169\pm 0.003\,\muas$.  The main error then comes from 
this measurement error, $\sigma(\ln\theta_*) = 0.11/2.43=0.045$.
This leads to $\mu_\rel = 1.16\pm 0.08\,\masyr$, which we argued
in Section~\ref{sec:binary-kb211105} is highly unlikely.

Finally, the blended light lies on the foreground main sequence of the CMD
in Figure~\ref{fig:allcmd}.  In principle, it might therefore be the
lens or a companion to lens.  We therefore carefully investigate the
offset between the magnified source and the baseline object
$\Delta\btheta(N,E) = \btheta_{\rm base}- \btheta_S$.  We make
four measurements by applying two independent algorithms (pyDIA and pySIS)
to two independent data sets (KMTC and KMTS).  The best-fit
values of the measurements are (in mas),
$\Delta\btheta_{\rm KMTC,pyDIA} = (+80,+72)$,
$\Delta\btheta_{\rm KMTS,pyDIA} = (+59,+24)$,
$\Delta\btheta_{\rm KMTC,pySIS} = (+61,+78)$, and
$\Delta\btheta_{\rm KMTS,pySIS} = (+106,+142)$.  

Before investigating the issue of measurement errors, we note
that 3 of the 4 measurements lead to $|\Delta\btheta|\la 0.1^{\prime\prime}$.
The surface density of foreground-main-sequence stars that are brighter
than the blend is just $39\,{\rm arcmin^{-2}}$, implying that the
probability for a random field star to lie within $0.1^{\prime\prime}$ is
just $p=3.4\times 10^{-4}$.  Thus, we must seriously consider the
possibility that the apparent offset is due to measurement error.  

The offset measurements have two sources of error: the error
in the source position (derived from difference images) and
the error in the baseline-object position (derived from
point-spread-function (PSF) photometry/astrometry of the baseline images).
We expect the first of these to be small because the difference
images are virtually free of systematic structures, apart from
the magnified source.  For example, in the two pySIS analyses,
we find standard deviations from the 6 magnified images to be
(again in mas)
$\sigma_{\rm KMTC} = (21,22)$ and
$\sigma_{\rm KMTS} = (50,25)$.  The scatter is substantially
smaller than the offsets, and it is plausible to treat these
6 measurements as independent, in which case the standard errors
of the mean are substantially smaller yet.

However, it is substantially more difficult to estimate the errors
in the DoPhot \citep{dophot} PSF-photometry measurement of the
baseline-object position.  While the baseline object appears
isolated on the image, and the nearest neighbor in the star catalog is
separated from the baseline object by $1.1^{\prime\prime}$, the astrometry
of the baseline object could easily be corrupted by a faint field star.
For example, an $I=20.75$ star separated by $0.5^{\prime\prime}$ would
not be separately resolved and would generate $0.1^{\prime\prime}$ error
in the measured position.  The surface density of such stars (without
accounting for incompleteness at these faint magnitudes) is 
$608\,{\rm arcmin^{-2}}$, implying that the expected number within
$0.5^{\prime\prime}$ is $p=13\%$.  Therefore, it is quite plausible
that the blended light is primarily due to the lens or a companion to the
lens, although the evidence in favor of this scenario is certainly not
definitive. We discuss the implications of this further in 
Section~\ref{sec:phys-kb211105}.

\section{{Physical Parameters}
\label{sec:phys}}

None of the four planets have sufficient information to precisely
specify the host mass and distance.  Moreover, several have multiple
solutions with significantly different mass ratios $q$ and/or different 
Einstein radii $\theta_\e$.  For any given solution, we can incorporate
Galactic-model priors into standard Bayesian techniques to obtain estimates
of the host mass $M_{\rm host}$ and distance $D_L$, 
as well as the planet mass $M_{\rm planet}$ 
and planet-host projected separation $a_\perp$.
See \citet{ob180567} for a description of the Galactic model and Bayesian
techniques.  However, in most cases we still have to decide how to
combine these separate estimates into a single ``quotable result''.
Moreover, in several cases, we also discuss how the nature of the planetary
systems can ultimately be resolved by future AO observations.  Hence,
we discuss each event separately below.

{\subsection{{KMT-2021-BLG-0712}\label{sec:phys-kb210712}}}

Because $\theta_\e$ and $\pi_\e$ are both measured, it might appear that we could
directly estimate the lens mass, $M=\theta_\e/\kappa\pi_\e$, and
distance, $D_L = \au(\pi_\e\theta_\e + \pi_S)$, where $\pi_S\sim 120\,\mas$.
However, because the two parallax solutions in Table~\ref{tab:0712parms}
differ significantly, this procedure yields two different pairs of values:
$(M/M_\odot,D_L/\kpc)=(0.18,2.7)$ and $(0.11,1.7)$.
Moreover, because the errors in $\pi_\e$ are not negligible, 
phase-space considerations will generally favor more distant lenses
within each solution.  And furthermore, they 
will also favor the $u_0>0$ solution due to its smaller $\pi_\e$.
Hence, in order to take account of phase space and properly weight these 
two solutions, it is essential conduct a Bayesian analysis.
As constraints, we include $t_\e$ (Table~\ref{tab:0712parms}), $\bpi_\e$
(Equations~(\ref{eqn:0712par+}) and (\ref{eqn:0712par-})) and $\theta_\e$
(Equations~(\ref{eqn:0712thetae_minus}) and (\ref{eqn:0712thetae_plus})).
Note that the parallax error ellipses can also be expressed in Equatorial
coordinates, in which case the errors in the cardinal directions are given
by Table~\ref{tab:0712parms} and the correlation coefficients are
$+0.53$ and $-0.73$ for the $u_0>0$ and $u_0<0$ solutions, respectively.
Formally, we also include the constraint on the lens flux $I_L>I_B=19.21$,
although as a practical matter it plays no role because the $\pi_\e$ and
$\theta_\e$ measurements already imply $I_L\ga 23$.

The results shown in Table~\ref{tab:physall} confirm the naive reasoning
given above.  First, the median lens distances are larger than the
naive estimates, while the 68\% confidence intervals are skewed toward
even larger distances.   Second, because $\theta_\e\equiv\sqrt{\kappa M\pi_\rel}$
is better constrained than $\pi_\e$,  the ``phase-space pressure'' toward
larger $D_L$ (smaller $\pi_\rel$), also pushes the host masses up relative
to the naive estimates and likewise causes them to be asymmetric toward
even higher values.  Third, the ``more populated'' regions of phase
space that are available to the $u_0>0$ solution due to its smaller $\pi_\e$,
gives it substantially higher Galactic-model weight, which more than
compensates for its slightly worse $\chi^2$.

The mass and distance distributions for the two solutions are shown
in the top two rows of Figure~\ref{fig:hist1}.

The planet is intermediate in mass between Neptune and Saturn and orbits
a mid-late M dwarf at about 3 kpc.

{\subsection{{KMT-2021-BLG-0909}\label{sec:phys-kb210909}}}

For this event, there is only one solution, for which there are two constraints:
$t_\e = 16.06\pm 0.73$ day (from Table~\ref{tab:0909parms}) and
$\theta_\e = 0.362\pm 0.049$ mas (from Equation~(\ref{eqn:0909thetae})).
The mass and distance distributions are presented
in the third row of Figure~\ref{fig:hist1}.  These show that the 
system most likely lies in the Galactic bulge.  

The planet is of Jovian mass and orbits
a middle M dwarf at about 6.5 kpc.

{\subsection{{KMT-2021-BLG-2478}\label{sec:phys-kb212478}}}

For this event, there are three well-understood constraints (on $t_\e$,
$\bpi_\e$, and $I_L$), and one constraint (on $\rho$) whose reliability
must still be investigated.  The first two constraints are given in 
Table~\ref{tab:2478parms}, with the correlation coefficients for $\bpi_\e$ 
being $-0.02$ and $+0.04$ for $u_0>0$ and $u_0<0$, 
respectively.  As discussed in 
Section~\ref{sec:cmd-kb212478}, the constraint on lens light is $I_L>I_B=18.15$.

As discussed in Sections~\ref{sec:anal-kb212478} and \ref{sec:cmd-kb212478},
the apparent constraints on $\rho$ from the light curve may be due to 
end-of-night systematics in the data, and they appear to be inconsistent with
other constraints.  We further investigate this by conducting Bayesian
analyses with and without the $\rho$ constraint, which we implement as
$\exp(-\chi^2(\rho)/2)$, where $\chi^2(\rho)$ is the envelope of $\chi^2$
with respect to $\rho$ (Figure~\ref{fig:2478envelope}) derived from the MCMC.

Table~\ref{tab:physall} shows the results of the Bayesian analysis with and
without the constraint on $\rho$.  It quantitatively
confirms the concerns based on the qualitative reasoning given above.
In particular, the ``Galactic weights'' are reduced by more than a factor
of 100 by the imposition of the $\rho$ constraint, implying that this
constraint is strongly inconsistent with the Bayesian priors from the
Galactic model.  When robust and well-understood measurements strongly conflict
with priors, it can be strong evidence of new discoveries.  However, in the
present case, these conflicts arise from a handful of data points
taken at high airmass.  Hence, the ``Adopted'' parameters in 
Table~\ref{tab:physall} reflect the Bayesian analysis that suppresses the
$\rho$ constraint.

The mass and distance distributions (bottom 4 rows of Figure~\ref{fig:hist1})
give another perspective on these same issues.  In particular, the bottom
two rows show that the $\rho$ constraint restricts the lens to being
within $D_L<2\,\kpc$ of the Sun, which has very little phase space.  Note,
however, that with or without this constraint, bulge lenses are virtually
excluded.

The planet is of Jovian mass and orbits
a late M dwarf at about 2.5 kpc.

{\subsection{{KMT-2021-BLG-1105}\label{sec:phys-kb211105}}}

For this event there are two well-established constraints (on $t_\e$ and
$\theta_\e$) as well as one potential constraint (on $I_L$)
that remains to be investigated.  The first, from Table~\ref{tab:1105parms},
is $t_\e=35.0\pm 1.9\,$day (or $t_\e=35.4\pm 1.9\,$day).  The second,
as discussed in Section~\ref{sec:cmd-kb211105}, is implemented
via a $\chi^2(\rho)$ envelope function (Figure~\ref{fig:1105envelope}) 
together with an estimate for
each simulated event with Einstein radius $\theta_{\e,i}$, of
$\rho_i = \theta_*/\theta_{\e,i}$, with $\theta_*$ given by Table~\ref{tab:cmd}.

The potential constraint on $I_L$ comes from the limit 
$I_L\geq I_{B,\rm cal}$, where we calibrate the blend as
$I_{B,\rm cal} = I_B  - I_{\rm cl} + I_{\rm cl,0} + A_I =18.86$,
with $I_B=19.00$ and $I_{\rm cl}- I_{\rm cl,0}= 2.88$ coming from
Table~\ref{tab:cmd} and  $A_I=2.75$ coming from the KMT webpage.
The reason that this constraint is ``potential'' is that our investigation
in Section~\ref{sec:cmd-kb211105} showed that the lens could be the
origin of this blended light, in which case using it as a limit
to exclude simulated events would bias the result toward lower masses.
Thus, to check whether the ``blend = lens'' scenario is plausible
within a Bayesian context, we first carry out the Bayesian analysis
both with and without this constraint.

The results are given in Table~\ref{tab:physall} and illustrated in
Figure~\ref{fig:hist2}.  They show that
the Bayesian estimates hardly change between the two cases.
However, they also show that roughly
5\% of the Galactic weight is eliminated by the flux constraint.
This indicates that ``blend = lens'' hypothesis is consistent with the
Bayesian priors at about the $2\,\sigma$ level (without yet taking into
consideration the low probability of finding a random field star 
$\la 0.1^{\prime\prime}$ of the event).  

Therefore, we
also conduct an additional Bayesian simulation under the constraint
$|I_L - I_B| < 0.2$.  Note that the width of this interval is somewhat
arbitrary: we just seek to distinguish simulated events that are roughly
consistent with the ``blend = lens'' hypothesis from those that are not.
The first point to note is that the host is a roughly solar mass star at
$D_L\sim 4\,\kpc$.  That is, it is substantially more massive than
the unconstrained estimate, but roughly at the same distance.  Second,
at this distance, the lens system is located about 200 pc above the Galactic
plane, and it therefore lies behind almost all the dust.  Because
$[(V-I)_B - (V-I)_{\rm cl} = -0.46$ (see Figure~\ref{fig:allcmd}), this
implies $(V-I)_{B,0}\simeq 0.60$, which is a very plausible value for
a solar-mass (or slightly more massive) star.

Thus, we find that the ``blend = lens'' scenario is consistent with all
the available constraints.  On the other hand, the hypothesis that
the blend is a companion to the lens is also plausible.  In this
case, the $\sim 0.1^{\prime\prime}$ astrometric offset would be explained
by the companion being roughly 400 au from the host (rather than
corruption of the astrometry by a faint field star).

We conclude that the blend is very likely to be either the host or a companion
to the host.  These scenarios could easily be distinguished by high 
resolution imaging by either AO on 8m class telescopes or the 
{\it Hubble Space Telescope (HST)}.  That is, a bright companion at
$\sim 0.1^{\prime\prime}$ would easily be detected, which would verify
the ``blend = companion'' scenario, while a contaminating random field star
at several tenths of an arcsecond (verifying the ``blend = lens'' scenario)
would be even more easily resolved.  However, pending such a resolution,
we advocate using the ``no $I$ constraint'' Bayesian analysis, which we
treat as ``Adopted'' in Table~\ref{tab:physall}.

If future high-resolution imaging (which could be done immediately) confirm
the ``blend = lens'' hypothesis, then KMT-2021-BLG-1105Lb would be one of the
rare microlensing planets that could be further studied using the
radial velocity (RV) technique.  If we adopt $M\simeq 1.1\,M_\odot$, which
is consistent with both Table~\ref{tab:physall} and $(V-I)_{B,0}=0.60$, and
we adopt $\pi_\rel \sim 0.13\,\mas$ (consistent with Table~\ref{tab:physall}),
then $\theta_\e \equiv \sqrt{\kappa M \pi_\rel}\simeq 1.1\,\mas$, and 
$a_\perp \sim 5.3\,\au$ (or 4.1 au).  
Considering that the semi-major axis is likely to be
larger than the projected separation by a factor $\sim 1.5^{1/2}=1.22$,
the orbital periods for Locals 1 and 2 are likely of order 16 and 11
years, respectively, with RV amplitudes of order 
$v\sin i \sim (25\,{\rm m\,s^{-1}})\sin i$ and 
$\sim (30\,{\rm m\,s^{-1}})\sin i$.  Adopting $A_I/E(V-I)\sim 1.3$ from
Figure~6 of \citet{nataf13}, we obtain 
$V_{B,\rm cal} = I_{B,\rm cal} + (V-I)_0 + A_I/1.3 \simeq 21.6$.  Hence,
assuming that future high-resolution imaging confirms that the blend is
the lens, it will be feasible to carry out RV measurements of the
requisite precision on this $(I,V)\simeq (18.9,21.6)$ star
on 30m class telescopes.

\section{{Discussion}
\label{sec:discuss}}

This is the second in a series of papers that aims to publish
all planets (and possible planets) that are detected by eye from
2021 KMTNet data and that are not published for other reasons.
Together, we have presented a total of 8 such planets.  In the course of
these efforts, we identified a total of 7 other events that warranted
detailed investigation but did not yield good planetary (or possibly planetary)
solutions.  Among the total of 15 events that were analyzed to prepare 
these two papers, none were possibly planetary but ultimately ambiguous.  
We have summarized that 8 other by-eye KMTNet planets have been published
of which 7 will likely enter the AnomalyFinder statistical sample, as well as
one possible planet.  Thus, to date, there are a total of 16 planets
from 2021 that are seemingly suitable
for statistical analysis.  These work-in-progress figures can be compared
to 2018, which is the only year with a complete sample of KMTNet planets,
as cataloged by \citet{2018prime} and \citet{2018subprime}.
In that case, of the 33 planets found by AnomalyFinder that were
suitable for statistical studies, 22 were discovered by eye, while
of the 8 possible planets, 3 were discovered by eye.  That is, there
were 22/3 discoveries for the full 2018 year compared to 16/1 discoveries
for the partial 2021 year.  Hence, this ongoing work is broadly consistent
with the only previous comprehensive sample.

\acknowledgments 
This research has made use of the KMTNet system operated by the Korea
Astronomy and Space Science Institute (KASI) and the data were obtained at
three host sites of CTIO in Chile, SAAO in South Africa, and SSO in
Australia.
This research was supported by the Korea Astronomy and Space Science
Institute under the R\&D program(Project No. 2022-1-830-04) supervised
by the Ministry of Science and ICT.
Work by C.H. was supported by the grant (2017R1A4A101517) of
National Research Foundation of Korea.
Work by H.Y. and W.Z. were partly supported by the National Science Foundation of China (Grant No. 12133005). 
J.C.Y. acknowledges support from U.S.\ N.S.F. Grant No. AST-2108414.
%
%

%

\input tabnames

\input tab0712

\input tab0909

\input tab2478

\input tab1105

\input tab1105_1L2S

\input tabcmd

\input tab_physall

\clearpage

\begin{figure}
\plotone{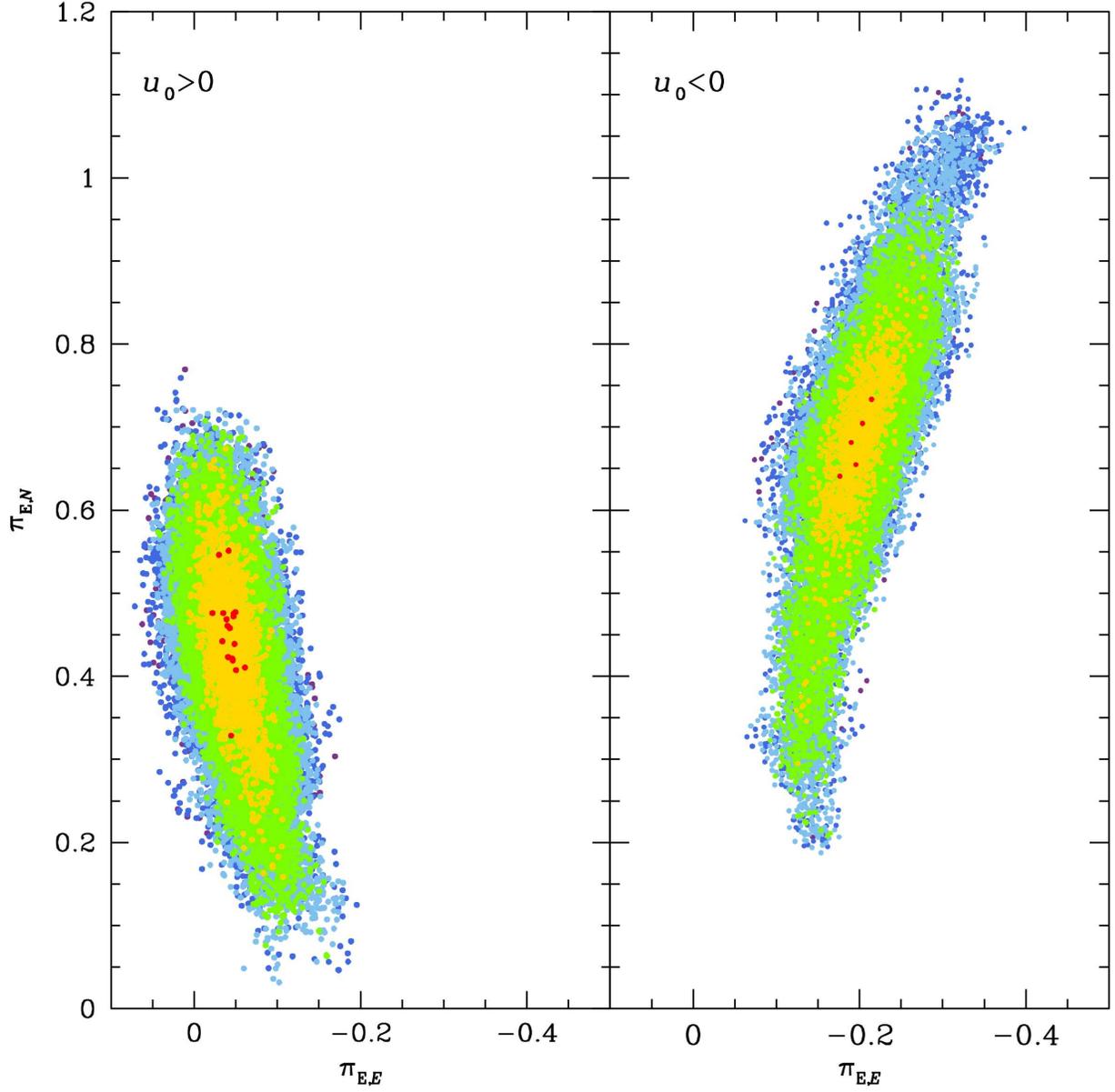}
\caption{Scatter plot of MCMC trials in the $\bpi_\e = (\pi_{\e,N},\pi_{\e,E})$
plane for the $u_0>0$ (left) and $u_0<0$ (right) solutions of 
KMT-2021-BLG-0712.  Points are colored (red, yellow, green, cyan, blue) 
if they are $\Delta\chi^2<(1,4,9,16,25)$ from the minimum.
}
\label{fig:0712par}
\end{figure}

\begin{figure}
\plotone{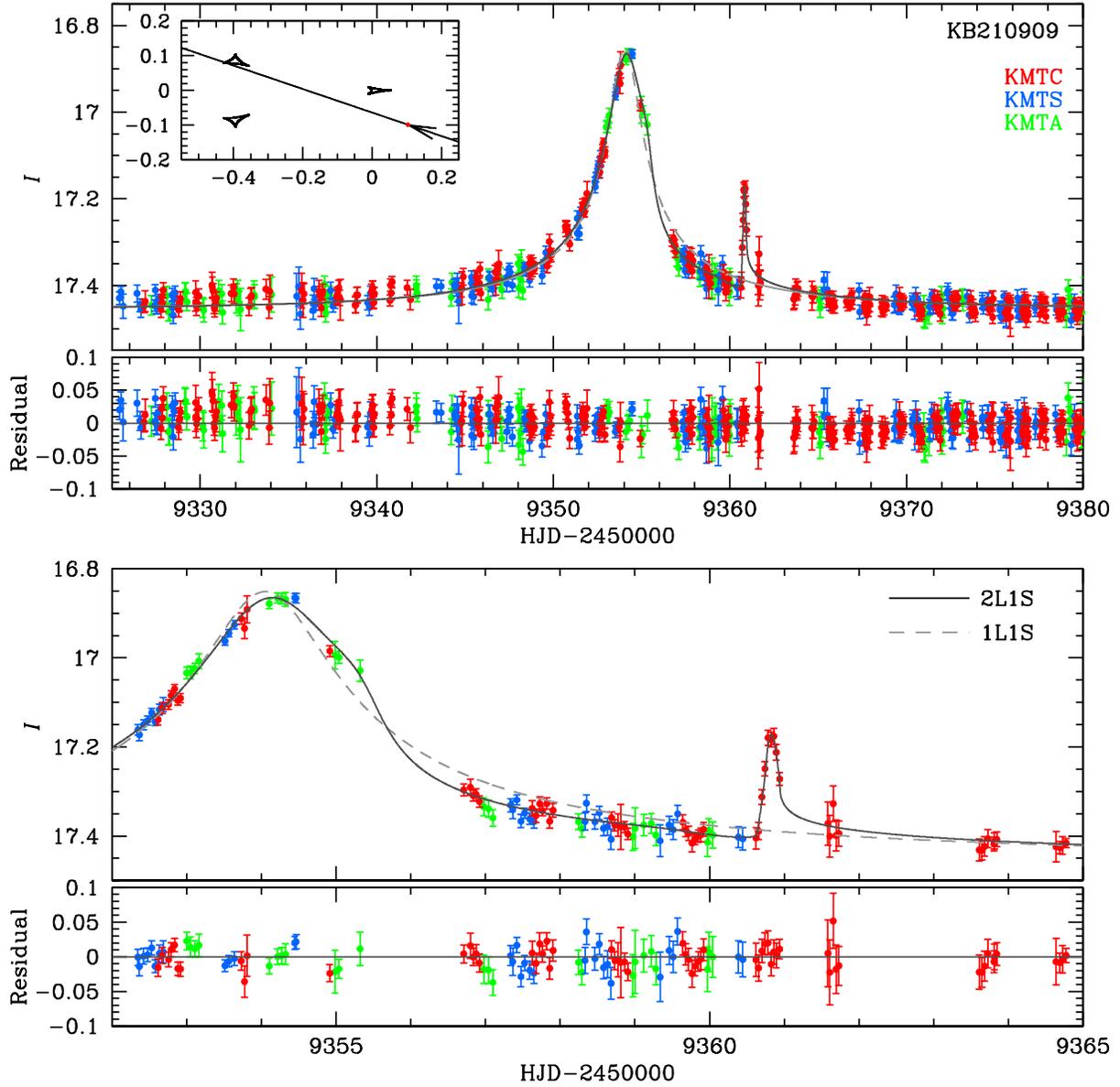}
\caption{Light-curve data and models for KMT-2021-BLG-0909.
Similar to Figure~\ref{fig:0712lc}, except that the 1L1S model is
also shown for comparison.
}
\label{fig:0909lc}
\end{figure}

\begin{figure}
\plotone{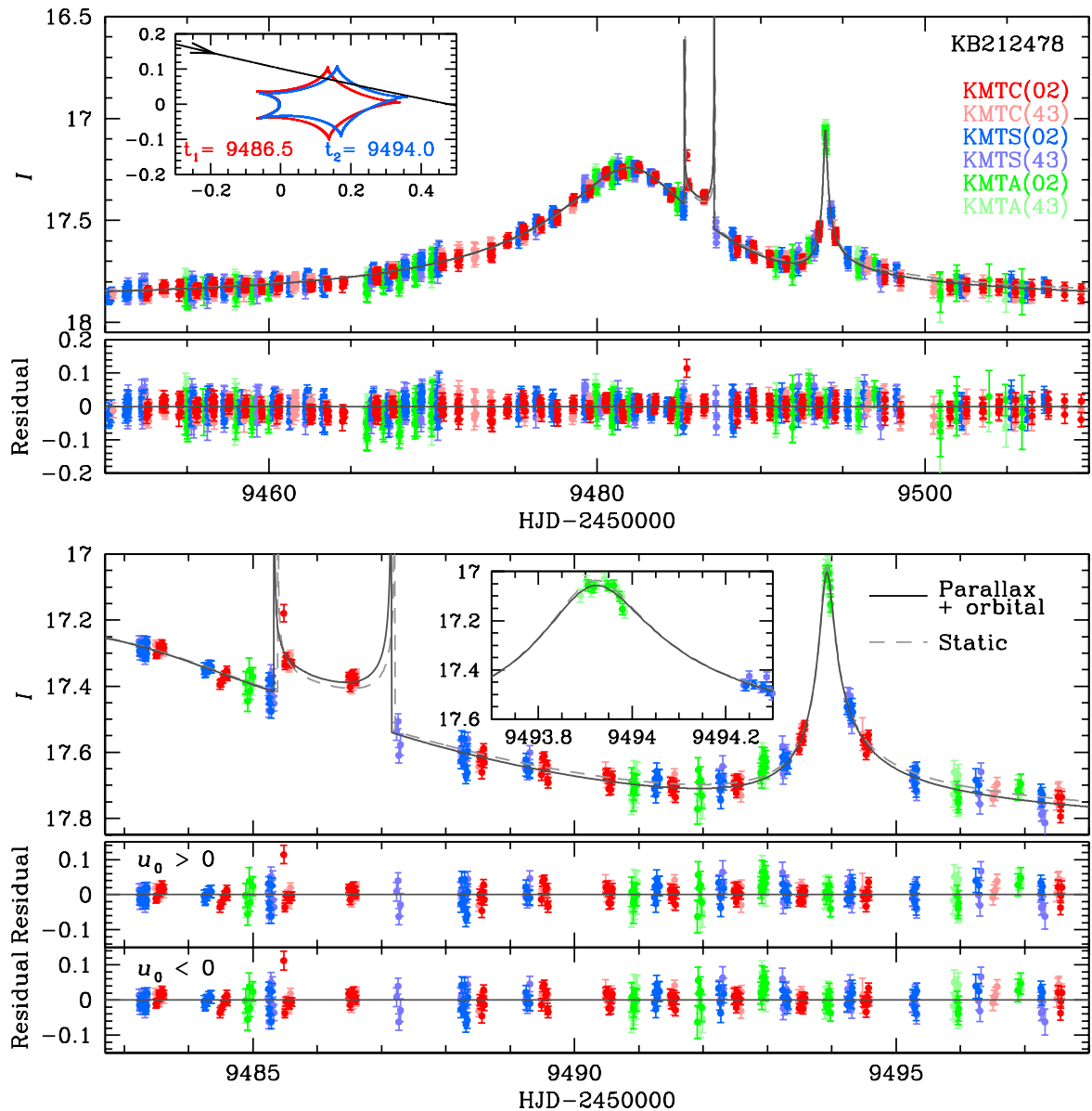}
\caption{Light-curve data and models for KMT-2021-BLG-2478.
Similar to Figure~\ref{fig:0712lc} except that the source-trajectory inset 
shows the caustic structure at two epochs and both the static and parallax
models are indicated.  In addition there is an inset in the lower
panel that highlight the KMTA coverage of the peak of the ``spike''.
}
\label{fig:2478lc}
\end{figure}

\begin{figure}
\plotone{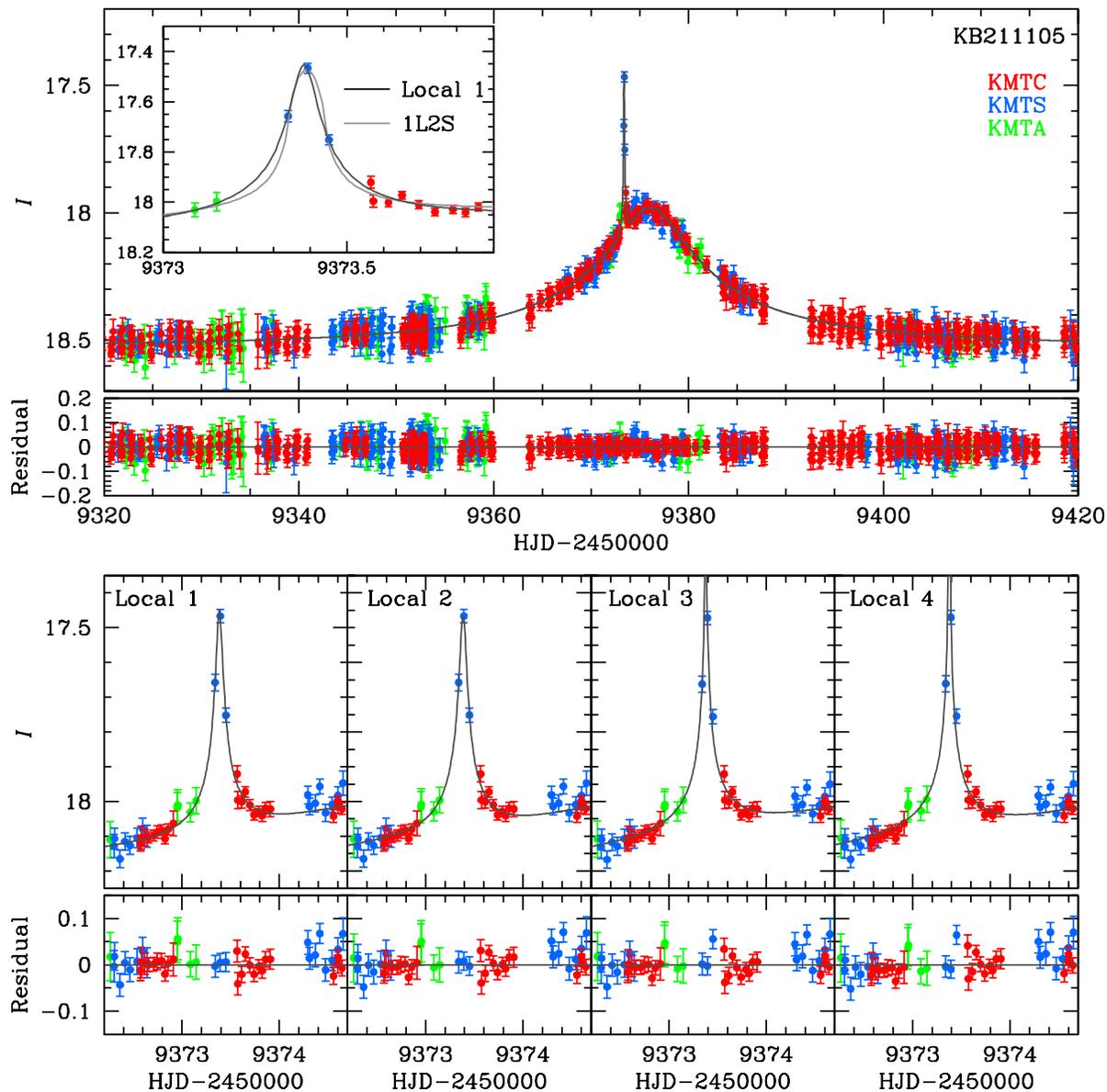}
\caption{Light-curve data and models for KMT-2021-BLG-1105.
Similar to Figure~\ref{fig:0712lc}, except that the anomaly region
is shown in separate panels for the four solutions at the bottom, while the
inset at the top compares the 1L2S solution to the Local-1 2L1S solution.
In addition, because there are four caustic topologies, these are shown
separately in Figure~\ref{fig:1105caust}.
}
\label{fig:1105lc}
\end{figure}

\begin{figure}
\plotone{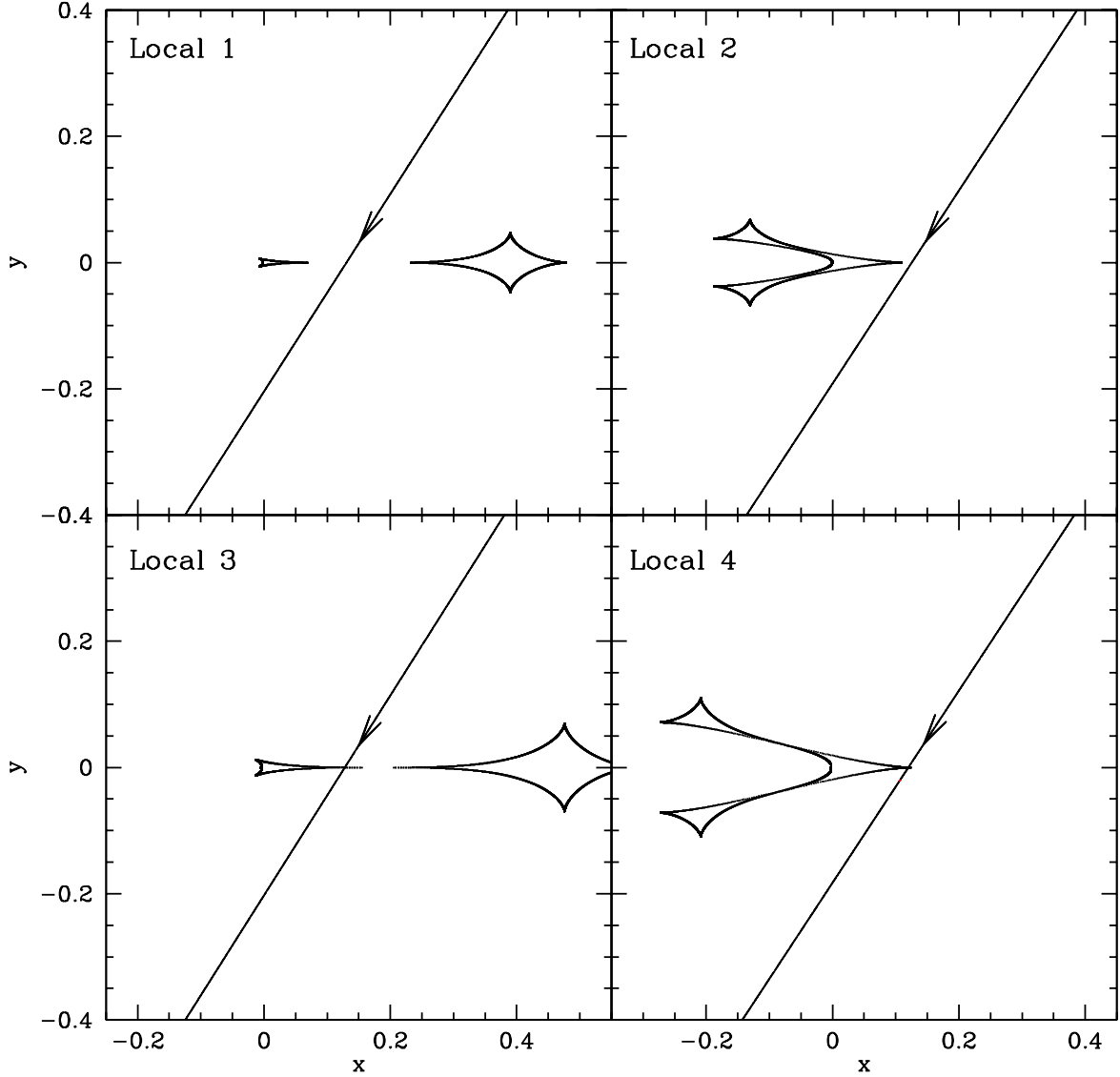}
\caption{Caustic topologies and source trajectories for each of the
four solutions of KMT-2021-BLG-1105.  Locals 1 and 2 constitute
an ``inner/outer'' degenerate pair satisfying 
$s^\dagger_+ \simeq s^\dagger\equiv \sqrt{s_{\rm inner}s_{\rm outer}}$ 
to high precision (Equation~(\ref{eqn:kb1105_heur})), while
Locals 3 and 4 constitute a second such pair that satisfy this
equation equally well.  The first pair are ridge-crossing, while the
second pair are cusp crossing.  The first pair are favored by $\Delta\chi^2>10$
and hence are adopted here.
}
\label{fig:1105caust}
\end{figure}

\begin{figure}
\plotone{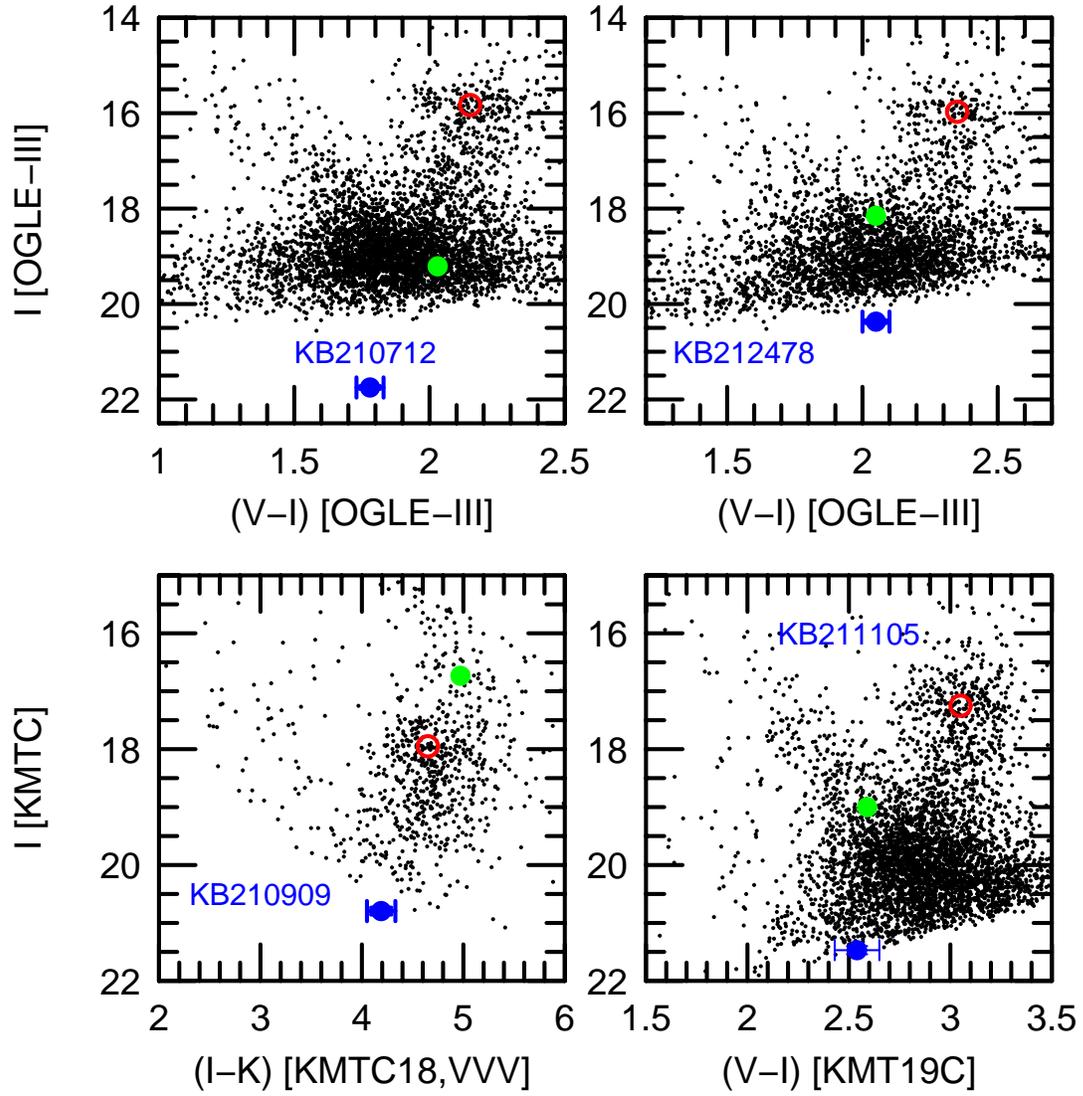}
\caption{CMDs for each of the four planets reported here.
The source positions (blue) and clump-giant centroids (red) are shown
for all events.  Where relevant, the blended light is shown in green.
}
\label{fig:allcmd}
\end{figure}

\begin{figure}
\plotone{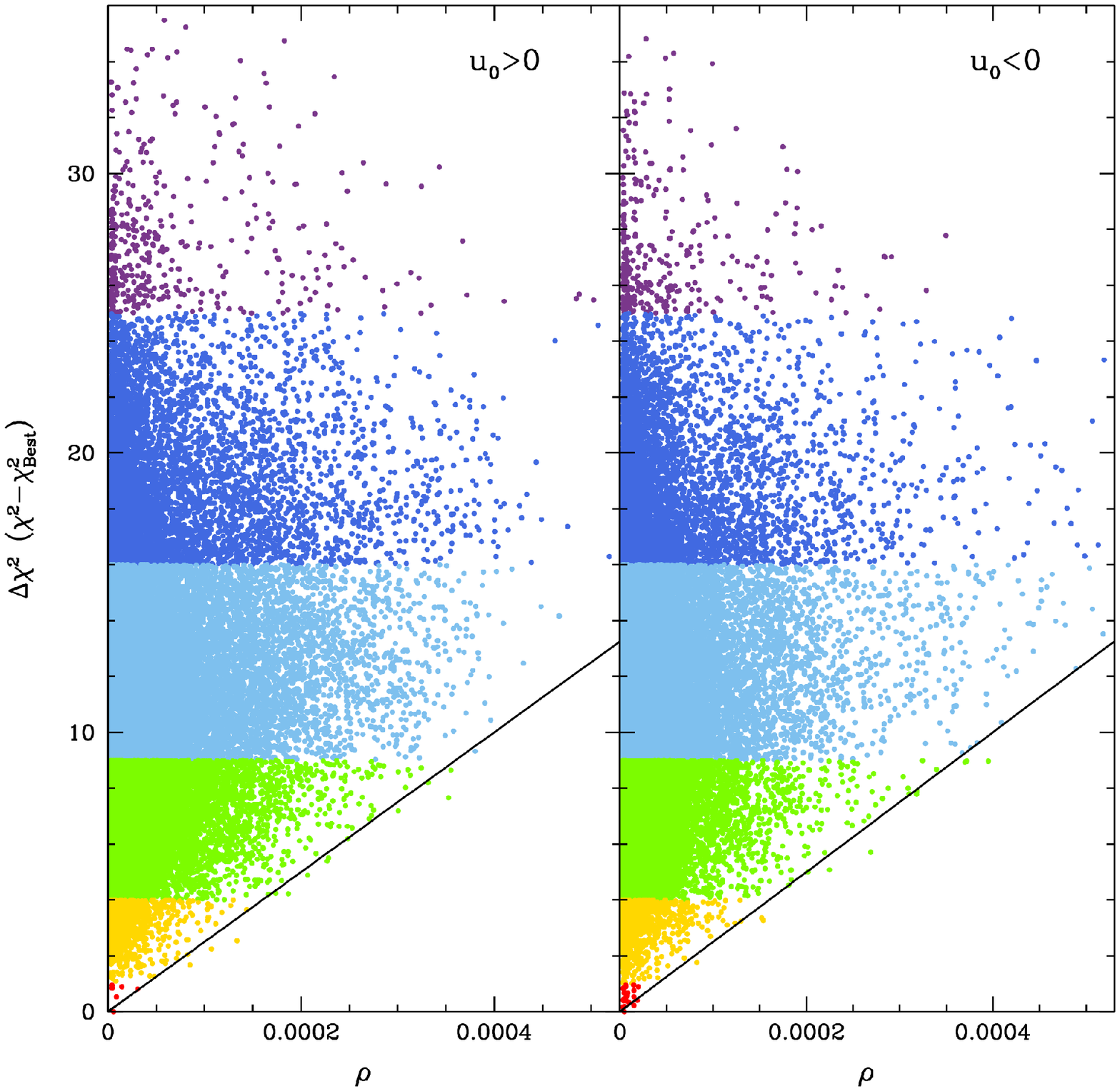}
\caption{Envelope function (solid curve) of $\Delta\chi^2(\rho)$ derived
from the lower limit of MCMC trials (colored points) for KMT-2021-BLG-2478.}
\label{fig:2478envelope}
\end{figure}

\begin{figure}
\plotone{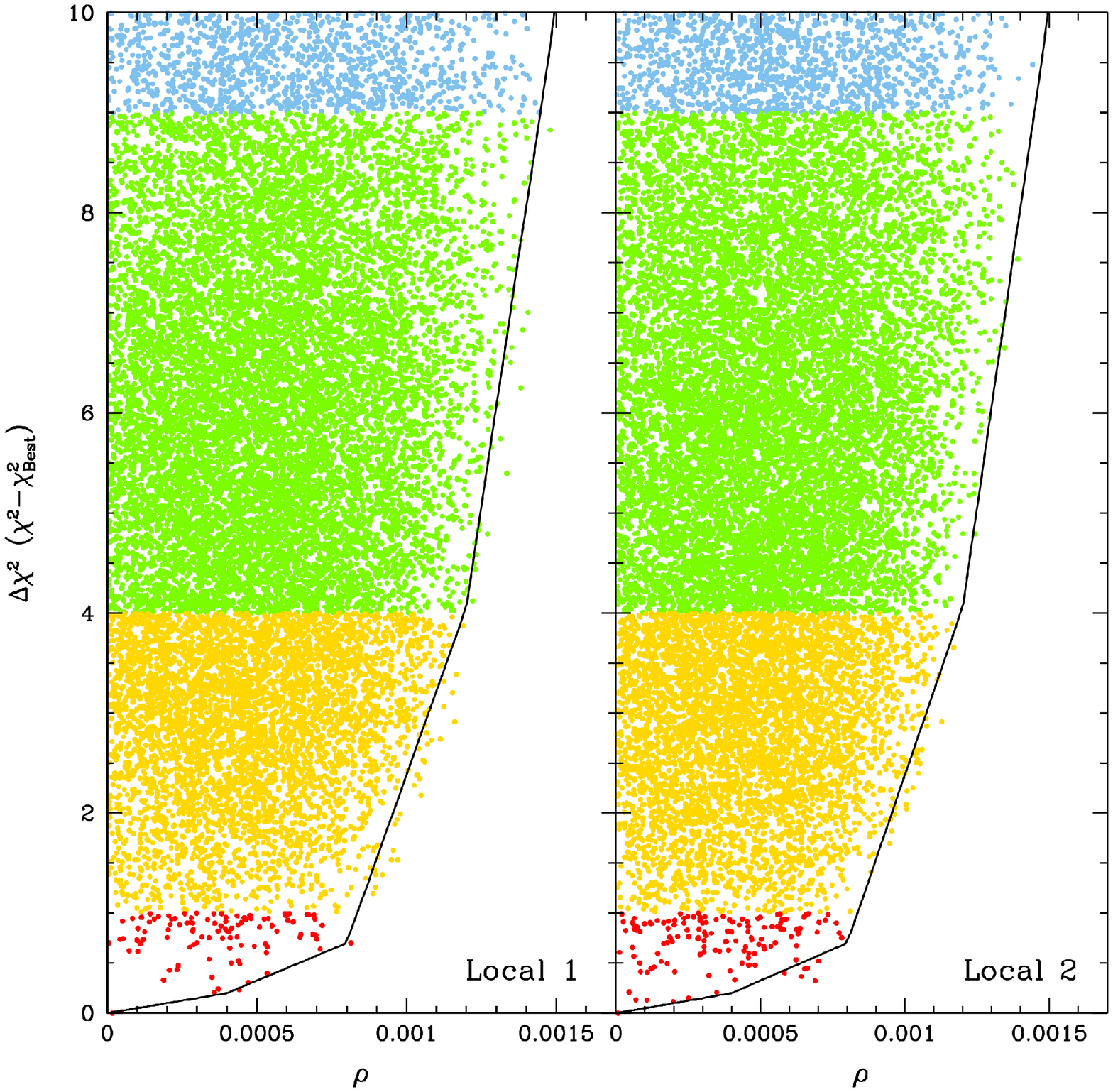}
\caption{Envelope function (solid curve) of $\Delta\chi^2(\rho)$ derived
from the lower limit of MCMC trials (colored points) for KMT-2021-BLG-1105.}
\label{fig:1105envelope}
\end{figure}

\begin{figure}
\plotone{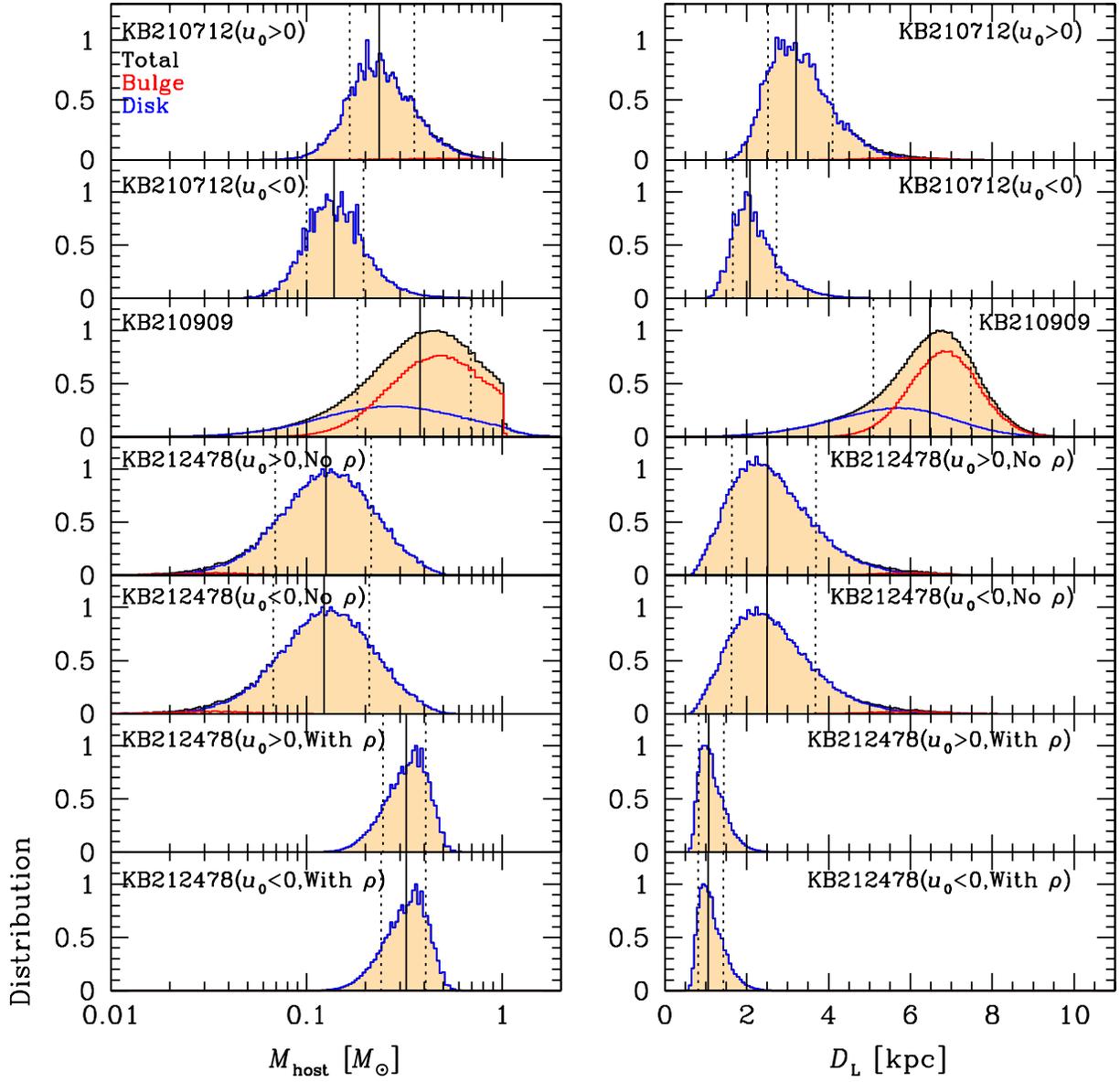}
\caption{Host-mass and system-distance distributions from Bayesian
analyses of various models (after imposing various constraints)
for three of the events analyzed in this
paper, KMT-2021-BLG-0712, KMT-2021-BLG-0909, and KMT-2021-BLG-2478.
For the first and and last of these events, there are two models
that are consistent with the light-curve data.  For KMT-2021-BLG-2478,
we consider two sets of constraints as described in 
Section~\ref{sec:phys-kb212478}.  Bulge-lens and disk-lens distributions 
are shown in red and blue, respectively, while their total is shown in
black.
}
\label{fig:hist1}
\end{figure}

\begin{figure}
\plotone{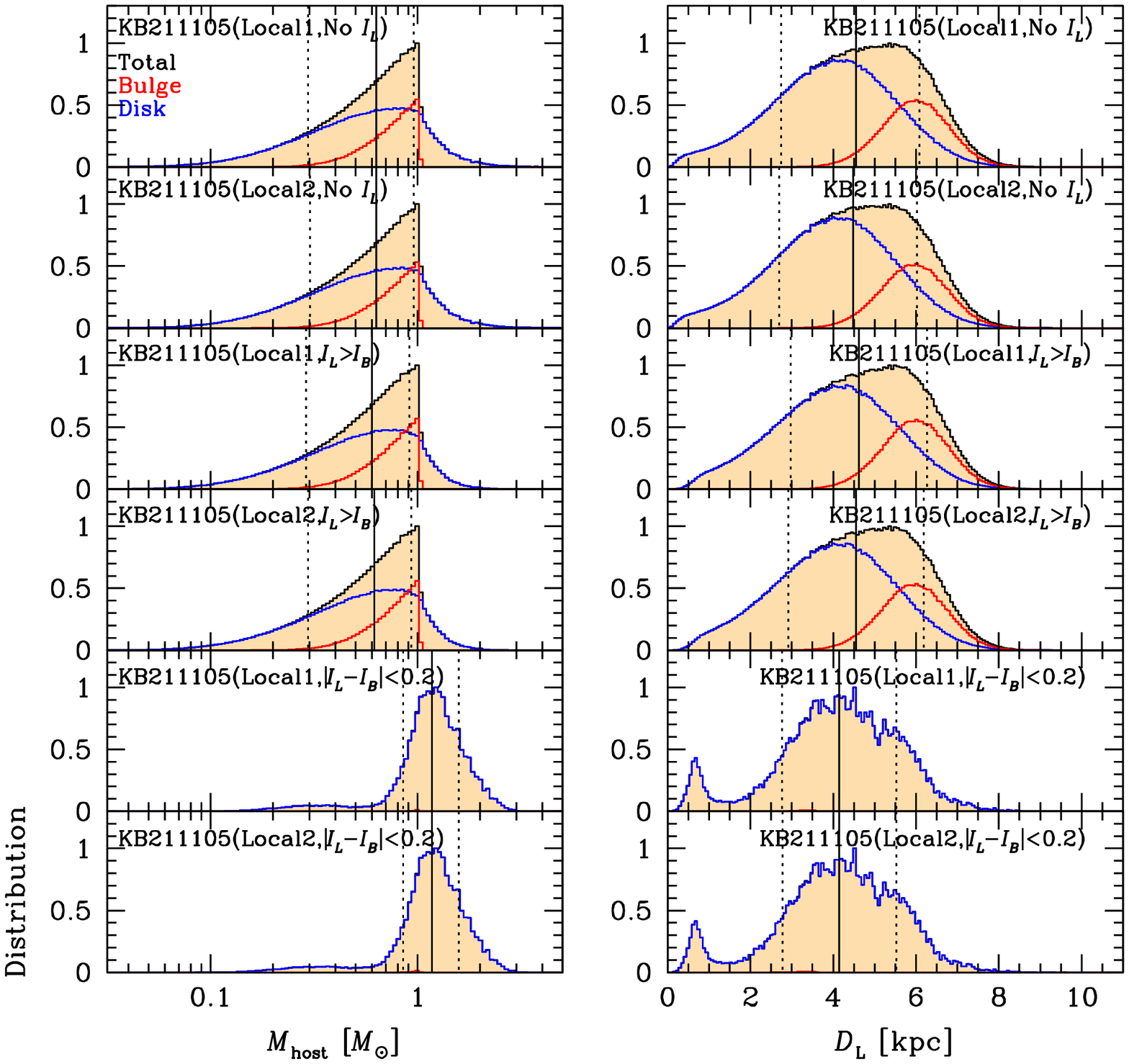}
\caption{Host-mass and system-distance distributions from Bayesian
analyses of two models (after imposing three different sets of constraints)
for KMT-2021-BLG-1105.  Color-coding is the same as Figure~\ref{fig:hist1}.
}
\label{fig:hist2}
\end{figure}

\end{document}

%% file: author.tex
\author{\textsc{
Yoon-Hyun Ryu$^{1}$, 
In-Gu Shin$^{2}$, 
Hongjing Yang$^{3}$, 
Andrew Gould$^{4,5}$, 
Michael D. Albrow$^{6}$, 
Sun-Ju Chung$^{1,7}$, 
Cheongho Han$^{8}$, 
Kyu-Ha Hwang$^{1}$, 
Youn Kil Jung$^{1}$, 
Yossi Shvartzvald$^{9}$, 
Jennifer C. Yee$^{2}$, 
Weicheng Zang$^{3}$,
Sang-Mok Cha$^{1,10}$, 
Dong-Jin Kim$^{1}$,
Seung-Lee Kim$^{1,7}$, 
Chung-Uk Lee$^{1,7}$, 
Dong-Joo Lee$^{1}$,
Yongseok Lee$^{1,10}$, 
Byeong-Gon Park$^{1,7}$, 
Richard W. Pogge$^{5}$
} }

\affil{$^{1}$Korea Astronomy and Space Science Institute, Daejon
34055, Republic of Korea}

\affil{$^{2}$ Center for Astrophysics $|$ Harvard \& Smithsonian, 60 Garden
St., Cambridge, MA 02138, USA}

\affil{$^{3}$ Department of Astronomy and Tsinghua Centre for Astrophysics, 
Tsinghua University, Beijing 100084, China}

\affil{$^{4}$Max-Planck-Institute for Astronomy, K\"{o}nigstuhl 17,
69117 Heidelberg, Germany}

\affil{$^{5}$Department of Astronomy, Ohio State University, 140 W.
18th Ave., Columbus, OH 43210, USA}

\affil{$^{6}$University of Canterbury, Department of Physics and
Astronomy, Private Bag 4800, Christchurch 8020, New Zealand}

\affil{$^{7}$Korea University of Science and Technology, Korea, 
(UST), 217 Gajeong-ro, Yuseong-gu, Daejeon, 34113, Republic of Korea}

\affil{$^{8}$Department of Physics, Chungbuk National University,
Cheongju 28644, Republic of Korea}

\affil{$^{9}$Department of Particle Physics and Astrophysics, 
Weizmann Institute of Science, Rehovot 76100, Israel}

\affil{$^{10}$School of Space Research, Kyung Hee University,
Yongin, Kyeonggi 17104, Republic of Korea}





%% file: tabnames.tex
 \begin{deluxetable}{lrrrrrr}
 \tablecolumns{7} \tablewidth{0pc}
 \tablecaption{\textsc{Event Names, Cadences, Alerts, and Locations}}
 \tablehead{\colhead{Name} & 
\colhead{$\Gamma\,({\rm hr}^{-1})$} &
\colhead{Alert Date} &
\colhead{RA$_{\rm J2000}$} &
\colhead{Dec$_{\rm J2000}$} &
\colhead{$l$} &
\colhead{$b$} }
 \startdata
KMT-2021-BLG-0712 & 4.0 &  1 May 2021 & 17:57:08.56 & $-$31:11:04.09 & $-0.66$ & $-3.29$\\
KMT-2021-BLG-0909 & 1.0 & 19 May 2021 & 17:42:28.85 & $-$27:38:20.90 & $+0.75$ & $+1.27$\\
KMT-2021-BLG-2478 & 4.0 & 14 Sep 2021 & 17:57:14.21 & $-29$:06:07.09 & $+1.16$ & $-2.27$\\
KMT-2021-BLG-1105 & 1.0 &  2 Jun 2021 & 17:42:55.74 & $-25$:30:32.08 & $+2.61$ & $+2.30$\\
\hline
 \enddata
 \tablecomments{The coordinates given here are for the nearest catalog 
stars (i.e., baseline objects). In Section~\ref{sec:cmd}
we discuss the offsets from these locations of the actual events.}
 \label{tab:names}
 \end{deluxetable}

%% file: tab0712.tex
\begin{deluxetable}{lccc}
	\tablecolumns{4} \tablewidth{0pc} \tablecaption{\textsc{Microlens Parameters for KMT-2021-BLG-0712}} \tablehead{ \colhead{ } & \colhead{
		}&
		\multicolumn{2}{c}{Parallax models} \\
		\cline{3-4} \colhead{Parameters } & \colhead{Standard}&
		\colhead{$u_0>0$}&\colhead{$u_0<0$} } \startdata
  $\chi^2/\rm{dof}$             &4263.666/4264       &4224.244/4260              &4222.276/4260       \\
  $t_0-2459340$                 &9.317 $\pm$ 0.115   &9.954 $\pm$ 0.144          &10.068 $\pm$ 0.129  \\
  $u_0$                         &0.147 $\pm$ 0.002   &0.145 $\pm$ 0.002          &-0.144 $\pm$ 0.002  \\
  $t_{\rm E}$ $(\rm{days})$     &90.865 $\pm$ 0.951  &88.503 $\pm$ 1.442         &100.583 $\pm$ 2.039 \\
  $s$                           &1.196 $\pm$ 0.002   &$1.196_{-0.012}^{+0.016}$  &1.206 $\pm$ 0.020   \\
  $q$ $(10^{-4})$               &4.751 $\pm$ 0.141   &$5.121_{-0.654}^{+0.998}$  &5.653 $\pm$ 1.157   \\
  log $q$ (mean)                &-3.323 $\pm$ 0.014  &-3.285 $\pm$ 0.062         &-3.251 $\pm$ 0.079  \\
  $\alpha$ $(\rm{rad})$         &0.467 $\pm$ 0.004   &$0.452_{-0.034}^{+0.027}$  &5.788 $\pm$ 0.039   \\
  $\rho$ $(10^{-4})$            &3.903 $\pm$ 0.330   &4.010 $\pm$ 0.449          &4.221 $\pm$ 0.551   \\
  $\pi_{\rm{E},\it{N}}$         &-                   &0.425 $\pm$ 0.104          &0.701 $\pm$ 0.111   \\
  $\pi_{\rm{E},\it{E}}$         &-                   &-0.047 $\pm$ 0.034         &-0.204 $\pm$ 0.042  \\
  $ds/dt$ $(\rm{yr}^{-1})$      &-                   &$0.008_{-0.199}^{+0.152}$  &-0.149 $\pm$ 0.262  \\
  $d\alpha/dt$ $(\rm{yr}^{-1})$ &-                   &$-0.688_{-0.321}^{+0.415}$ &-0.902 $\pm$ 0.602  \\
  $I_S$ [KMTC01,pySIS]        &21.618 $\pm$ 0.013  &21.644 $\pm$ 0.017         &21.653 $\pm$ 0.018  \\
  $I_B$ [KMTC01,pySIS]        &18.676 $\pm$ 0.001  &18.672 $\pm$ 0.002         &18.673 $\pm$ 0.002  \\
  $t_*$ $(\rm{hours})$          &0.851 $\pm$ 0.069   &0.852 $\pm$ 0.092          &1.019 $\pm$ 0.138   \\
\enddata
\label{tab:0712parms}
\end{deluxetable}

%% file: tab0909.tex
\begin{deluxetable}{lcccc}
\tablecolumns{5} \tablewidth{0pc} \tablecaption{\textsc{Microlens Parameters for KMT-2021-BLG-0909}} \tablehead{\colhead{Parameters}
& \colhead{2L1S}  } \startdata
  $\chi^2/\rm{dof}$             &1636.961/1637                 \\
  $t_0-2459350$                 &4.073 $\pm$ 0.017           \\
  $u_0$                         &0.060 $\pm$ 0.004           \\
  $t_{\rm E}$ $(\rm{days})$     &16.046 $\pm$ 0.733          \\
  $s$                           &0.823 $\pm$ 0.008           \\
  $q$ $(10^{-3})$               &3.174 $\pm$ 0.446   \\
  log $q$ (mean)                &-2.497 $\pm$ 0.063          \\
  $\alpha$ $(\rm{rad})$         &3.472 $\pm$ 0.012           \\
  $\rho$ $(10^{-3})$            &3.326 $\pm$ 0.308 \\
  $I_S$ [KMTC,pySIS]            &20.836 $\pm$ 0.060          \\
  $I_B$ [KMTC,pySIS]            &17.503 $\pm$ 0.003  \\
  $t_*$ $(\rm{hours})$          &1.284 $\pm$ 0.085   \\
\enddata
\label{tab:0909parms}
\end{deluxetable}

%% file: tab2478.tex
\begin{deluxetable}{lccc}
	\tablecolumns{4} \tablewidth{0pc} \tablecaption{\textsc{Microlens Parameters for KMT-2021-BLG-2478}} \tablehead{ \colhead{ } & \colhead{
		}&
		\multicolumn{2}{c}{Parallax + orbital motion models} \\
		\cline{3-4} \colhead{Parameters } & \colhead{Standard}&
		\colhead{$u_0>0$}&\colhead{$u_0<0$} } \startdata
  $\chi^2/\rm{dof}$             &7720.303/7578               &7573.830/7574              &7574.077/7574       \\
  $t_0-2459480$                 &2.167 $\pm$ 0.018           &2.243 $\pm$ 0.033          &2.244 $\pm$ 0.033  \\
  $u_0$                         &0.081 $\pm$ 0.001           &0.099 $\pm$ 0.002          &-0.099 $\pm$ 0.002  \\
  $t_{\rm E}$ $(\rm{days})$     &40.603 $\pm$ 0.402          &33.689 $\pm$ 0.518         &33.489 $\pm$ 0.511 \\
  $s$                           &1.058 $\pm$ 0.001           &1.060 $\pm$ 0.002          &1.060 $\pm$ 0.002   \\
  $q$ $(10^{-3})$               &3.740 $\pm$ 0.121           &6.087 $\pm$ 0.479          &6.134 $\pm$ 0.480   \\
  log $q$ (mean)                &-2.428 $\pm$ 0.015          &-2.217 $\pm$ 0.034         &-2.213 $\pm$ 0.034  \\
  $\alpha$ $(\rm{rad})$         &0.274 $\pm$ 0.001           &0.233 $\pm$ 0.014          &6.051 $\pm$ 0.014   \\
  $\rho$ $(10^{-5})$            &$2.403_{-1.736}^{+7.143}$   &$2.452_{-1.763}^{+7.291}$  &$1.915_{-1.392}^{+5.232}$\\
  $\pi_{\rm{E},\it{N}}$         &-                           &0.100 $\pm$ 0.209          &0.113 $\pm$ 0.212   \\
  $\pi_{\rm{E},\it{E}}$         &-                           &0.518 $\pm$ 0.064          &0.522 $\pm$ 0.066  \\
  $ds/dt$ $(\rm{yr}^{-1})$      &-                           &0.871 $\pm$ 0.165          &0.872 $\pm$ 0.169  \\
  $d\alpha/dt$ $(\rm{yr}^{-1})$ &-                           &1.285 $\pm$ 0.561          &-0.918 $\pm$ 0.572  \\
  $I_S$ [KMTC(01),pySIS]        &20.724 $\pm$ 0.013          &20.497 $\pm$ 0.019         &20.494 $\pm$ 0.019  \\
  $I_B$ [KMTC(01),pySIS]        &17.971 $\pm$ 0.001          &17.992 $\pm$ 0.002         &17.992 $\pm$ 0.002  \\
  $t_*$ $(\rm{hours})$          &$0.023_{-0.017}^{+0.070}$   &$0.020_{-0.014}^{+0.059}$  &$0.015_{-0.011}^{+0.042}$\\
  $\beta$ $(\theta_*\equiv0.4\muas)$   &-                     &$0.057_{-0.041}^{+0.145}$  &$0.028_{-0.020}^{+0.083}$\\
\enddata
\label{tab:2478parms}
\end{deluxetable}

%% file: tab1105.tex
\begin{deluxetable}{lcccc}
\tablecolumns{5} \tablewidth{0pc} \tablecaption{\textsc{Microlens Parameters for KMT-2021-BLG-1105}} \tablehead{\colhead{Parameters} & \colhead{Local 1} & \colhead{Local 2} & \colhead{Local 3} & \colhead{Local 4}} \startdata
  $\chi^2/\rm{dof}$             &1784.854/1785               &1788.955/1785               &1795.837/1785              &1806.507/1785               \\
  $t_0-2459370$                 &5.826 $\pm$ 0.036           &5.856 $\pm$ 0.036           &5.711 $\pm$ 0.036          &5.780 $\pm$ 0.035           \\
  $u_0$                         &0.109 $\pm$ 0.007           &0.107 $\pm$ 0.007           &0.107 $\pm$ 0.007          &0.101 $\pm$ 0.005           \\
  $t_{\rm E}$ $(\rm{days})$     &34.965 $\pm$ 1.919          &$35.374_{-1.739}^{+2.264}$  &34.425 $\pm$ 1.985         &$36.248_{-1.488}^{+1.887}$  \\
  $s$                           &1.214 $\pm$ 0.008           &0.939 $\pm$ 0.008           &1.265 $\pm$ 0.013          &0.899 $\pm$ 0.010           \\
  $q$ $(10^{-3})$               &1.984 $\pm$ 0.200           &1.934 $\pm$ 0.191           &4.939 $\pm$ 0.694          &4.573 $\pm$ 0.614           \\
  log $q$ (mean)                &-2.703 $\pm$ 0.044          &-2.714 $\pm$ 0.043          &-2.308 $\pm$ 0.062         &-2.341 $\pm$ 0.058          \\
  $\alpha$ $(\rm{rad})$         &2.140 $\pm$ 0.009           &2.148 $\pm$ 0.010           &2.136 $\pm$ 0.010          &2.152 $\pm$ 0.009           \\
 $\rho$ $(10^{-3})$            &$<1.3$  &$<1.3$  &$<1.3$  &$<1.3$  \\
  $I_S$ [KMTC,pySIS]            &21.233 $\pm$ 0.074          &$21.258_{-0.070}^{+0.085}$  &21.207 $\pm$ 0.080         &21.294 $\pm$ 0.062          \\
  $I_B$ [KMTC,pySIS]            &18.614 $\pm$ 0.006          &18.612 $\pm$ 0.006          &18.616 $\pm$ 0.006         &18.609 $\pm$ 0.004          \\
  $t_*$ $(\rm{hours})$          &$< 1.1$ &$< 1.1$ &$< 1.1$ &$< 1.1$ \\
\enddata
\label{tab:1105parms}
\end{deluxetable}

%% file: tab1105_1L2S.tex
\begin{deluxetable}{lccc}
\tablecolumns{2} \tablewidth{0pc} \tablecaption{\textsc{1L2S for KMT-2021-BLG-1105}} \tablehead{\colhead{Parameters} & \colhead{1L2S}& \colhead{$\rho_2=0.001$}& \colhead{$\rho_2=0.0$} } \startdata
  $\chi^2/\rm{dof}$             &1790.359/1785       &1800.137/1786               &1803.805/1786    \\
  $t_{0,1}-2459370$             &6.162 $\pm$ 0.041   &6.151 $\pm$ 0.041           &6.192 $\pm$ 0.043  \\
  $u_{0,1}$                     &0.105 $\pm$ 0.010   &$0.077_{-0.005}^{+0.007}$   &0.103 $\pm$ 0.010  \\
  $t_{\rm E}$ $(\rm{days})$     &37.787 $\pm$ 3.115  &49.293 $\pm$ 3.400          &38.450 $\pm$ 3.202  \\
  $t_{0,2}-2459370$             &3.389 $\pm$ 0.002   &$3.388_{-0.004}^{+0.003}$   &3.381 $\pm$ 0.004  \\
  $u_{0,2}$ $(10^{-3})$         &0.000 $\pm$ 0.313   &0.025 $\pm$ 0.322           &0.654 $\pm$ 0.101  \\
  $\rho_2$ $(10^{-3})$          &1.416 $\pm$ 0.139   &1                       &-                  \\
  $q_F$ $(10^{-2})$             &1.021 $\pm$ 0.055   &1.065 $\pm$ 0.072           &1.186 $\pm$ 0.088   \\
  $I_S$ [KMTC,pySIS]            &21.355 $\pm$ 0.113  &$21.703_{-0.098}^{+0.080}$  &21.381 $\pm$ 0.113  \\
  $I_B$ [KMTC,pySIS]            &18.605 $\pm$ 0.007  &$18.586_{-0.003}^{+0.005}$  &18.603 $\pm$ 0.007   \\
  $I_{S,2}$ [KMTC,pySIS]        &26.344 $\pm$ 0.103  &$26.657_{-0.106}^{+0.065}$  &26.211 $\pm$ 0.111 \\
  $t_{*,2}$ $(\rm{hours})$      &1.284 $\pm$ 0.068   &1.183 $\pm$ 0.082           &-                    \\
\enddata
\label{tab:1105_1L2S}
\end{deluxetable}

%% file: tabcmd.tex
\begin{deluxetable}{lrrrr}
\tablecolumns{6} \tablewidth{0pc}
\tablecaption{\textsc{CMD Parameters for Four 2021 Planets}}
\tablehead{\colhead{Parameter} &
\colhead{KB210712} &
\colhead{KB210909} &
\colhead{KB212478} &
\colhead{KB211105}}
\startdata
$(V-I)_{\rm s}$    & 1.78$\pm$0.05 & N.A.         & 2.05$\pm$0.05  & 2.50$\pm$0.08\\ 
$(V-I)_{\rm cl}$   & 2.15$\pm$0.03 & N.A.         & 2.35$\pm$0.03 & 3.05$\pm$0.04\\ 
$(V-I)_{\rm cl,0}$  & 1.06           & 1.06        & 1.06 & 1.06 \\
$(V-I)_{\rm s,0}$   & 0.69$\pm0.06$ & 0.81$\pm$0.08 & 0.76$\pm$0.06& 0.51$\pm$0.09 \\ 
$I_{\rm s}$        & 21.75$\pm$0.02 & 20.80$\pm$0.06 & 20.37$\pm$0.03 & 21.49$\pm$0.07\\ 
$I_{\rm cl}$       & 15.83$\pm$0.05 & 17.95$\pm$0.05 & 15.97$\pm$0.05 & 17.25$\pm$0.05\\ 
$I_{\rm cl,0}$     & 14.48          & 14.41         & 14.39          & 14.37\\ 
$I_{\rm s,0}$      & 20.40$\pm$0.06 & 17.26$\pm$0.08 & 18.97$\pm$0.06 & 18.61$\pm$0.09\\  
$\theta_*$ ($\muas$) & 0.255$\pm 0.023$ & 1.205$\pm$0.163 & 0.533$\pm$0.043 & 0.482$\pm$0.053\\ 
\enddata
\tablecomments{Event names are abbreviations, e.g., KMT-2021-BLG-0712.
$[(V-I),I]_S$ and $[(V-I),I]_{\rm cl}$ 
for KB210712 and KB212478 are based on calibrated OGLE-III
photometry, while the other two events have instrumental KMT photometry.
}
\label{tab:cmd}
\end{deluxetable}

%% file: tab_physall.tex
\begin{deluxetable}{lccccccc}
\tablecolumns{8} 
\tablewidth{0pc}\tablecaption{\textsc{Physical properties}} 
\tablehead{\colhead{Event} & \multicolumn{4}{c}{} & \colhead{} &
\multicolumn{2}{c}{Relative Weights}\\
\cline{7-8} \colhead{Models}&\multicolumn{4}{c}{Physical Properties}&  &
\colhead{Gal.Mod.} & \colhead{$\chi^2$}} \startdata
 KB210712  &$M_{\rm host}$ $[M_\sun]$  &$M_{\rm planet}$ $[M_\earth]$  &$D_{\rm L}$ [kpc] &$a_\bot$ [au]\\
 $u_0>0$ &$0.23_{-0.07}^{+0.12}$  &$39.90_{-11.90}^{+20.60}$ &$3.20_{-0.68}^{+0.91}$ &$2.29_{-0.42}^{+0.64}$ &&1.000 &0.374\\
 $u_0<0$ &$0.14_{-0.04}^{+0.06}$  &$25.76_{-7.11}^{+11.28}$  &$2.08_{-0.42}^{+0.65}$ &$1.57_{-0.26}^{+0.42}$ &&0.055 &1.000\\
 Adopted &$0.22_{-0.07}^{+0.12}$  &$38.19_{-12.24}^{+20.52}$ &$3.09_{-0.75}^{+0.94}$ &$2.22_{-0.48}^{+0.66}$ && &\\
 \cline{1-8}
 KB210909  &$M_{\rm host}$ $[M_\sun]$  &$M_{\rm planet}$ $[M_J]$  &$D_{\rm L}$ [kpc] &$a_\bot$ [au]\\
 $     $ &$0.38_{-0.20}^{+0.30}$  &$1.26_{-0.66}^{+1.01}$ &$6.48_{-1.38}^{+1.00}$ &1.75 $\pm$ 0.42 && &\\
 \cline{1-8}
 KB212478  &$M_{\rm host}$ $[M_\sun]$  &$M_{\rm planet}$ $[M_J]$  &$D_{\rm L}$ [kpc] &$a_\bot$ [au]\\
 No $\rho$ constraint\\
 $u_0>0$ &$0.12_{-0.06}^{+0.09}$  &$0.80_{-0.35}^{+0.56}$  &$2.51_{-0.87}^{+1.18}$ &1.38 $\pm$ 0.26 &&1.000 &1.000\\
 $u_0<0$ &$0.12_{-0.05}^{+0.09}$  &$0.79_{-0.35}^{+0.56}$  &$2.50_{-0.87}^{+1.18}$ &1.37 $\pm$ 0.26 &&0.987 &0.884\\
 With $\rho$ constraint\\
 $u_0>0$ &0.33 $\pm$ 0.08         &2.08 $\pm$ 0.52         &$1.06_{-0.24}^{+0.37}$ &1.64 $\pm$ 0.20 &&0.0008 &1.000\\
 $u_0<0$ &0.32 $\pm$ 0.08         &2.08 $\pm$ 0.52         &$1.06_{-0.24}^{+0.37}$ &1.64 $\pm$ 0.20 &&0.0008 &0.884\\
 Adopted &$0.12_{-0.06}^{+0.09}$  &$0.80_{-0.35}^{+0.56}$  &$2.50_{-0.87}^{+1.18}$ &1.38 $\pm$ 0.26 && &\\
\cline{1-8}
 KB211105  &$M_{\rm host}$ $[M_\sun]$  &$M_{\rm planet}$ $[M_J]$  &$D_{\rm L}$ [kpc] &$a_\bot$ [au]\\
 No $I_L$ constraint\\
 Local 1 &0.63 $\pm$ 0.33         &1.30 $\pm$ 0.68         &4.55 $\pm$ 1.67        &$3.64_{-1.19}^{+0.93}$ &&1.000 &1.000\\
 Local 2 &0.63 $\pm$ 0.33         &1.28 $\pm$ 0.67         &4.48 $\pm$ 1.66        &$2.82_{-0.93}^{+0.72}$ &&0.975 &0.129\\
 $I_L>I_B$ constraint\\
 Local 1 &0.61 $\pm$ 0.31         &1.27 $\pm$ 0.65         &4.62 $\pm$ 1.65        &$3.60_{-1.14}^{+0.89}$ &&0.946 &1.000\\
 Local 2 &0.61 $\pm$ 0.32         &1.24 $\pm$ 0.64         &4.55 $\pm$ 1.64        &$2.78_{-0.89}^{+0.69}$ &&0.920 &0.129\\
 $|I_L-I_B|<0.2$ const.\\
 Local 1 &$1.17_{-0.31}^{+0.42}$  &$2.43_{-0.64}^{+0.88}$  &4.14 $\pm$ 1.38        &4.79 $\pm$ 0.87        &&0.023 &1.000\\
 Local 2 &$1.17_{-0.31}^{+0.42}$  &$2.37_{-0.63}^{+0.85}$  &4.15 $\pm$ 1.38        &3.70 $\pm$ 0.67        &&0.024 &0.129\\
 Adopted &0.63 $\pm$ 0.33         &1.30 $\pm$ 0.68         &4.54 $\pm$ 1.67        &3.54 $\pm$ 1.06 \\

 \enddata
 \label{tab:physall}
\end{deluxetable}